\documentclass [twoside,12pt] {article}
\usepackage{setspace,graphicx,bm,fancyhdr}
\usepackage{amsmath,amsthm,amssymb,latexsym}
\usepackage{color}
\usepackage{graphicx}

\usepackage{authblk}

\newcommand{\chp}[1]{\mbox{$\stackrel{\wedge}{#1}$}}

\newcommand{\ch}[1]{\mbox{$\stackrel{\sim}{#1}$}}
\newcommand{\vct}[1]{\mbox{$\stackrel{\rightharpoonup}{#1}$}}

\newcommand{\slas}[1]{\mbox{${{#1} \!\!\! /}$}}

\begin{document}
\title{\bf Computation of the leading order contributions to the Lamb shift for the H atom using spectral regularization} 
\author{John Mashford \\
School of Mathematics and Statistics \\
University of Melbourne, Victoria 3010, Australia \\
E-mail: mashford@unimelb.edu.au}
\date{\today}
\maketitle

\begin{abstract}

The Uehling contribution to the Lamb shift can be computed exactly in terms of the Uehling potential function. However derivations of this function are complex involving avoiding divergences using intricate techniques from early quantum field theory (QFT) or else more modern approaches using charge and mass renormalization. In the present paper we derive the Uehling potential function in a fairly conceptually straightforward way not involving renormalization in which the vacuum polarization tensor is viewed as a Lorentz invariant 2-tensor valued measure on Minkowski space. Furthermore we compute a complex matrix valued potential function for the electron self-energy contribution to the Lamb shift. The resulting potential function is derived in a conceptually simple way not involving renormalization and can be used for higher order computations in QFT involving multiple loops.

\end{abstract}

\tableofcontents

\section{Introduction}

The Lamb shift is a phenomenon which is closely tied up with the instigation and development of quantum field theory (QFT). At the time of its discovery in 1947 by Lamb and Retherford \cite{60,51} it created considerable interest in the physics community and provided impetus for the development of the renormalization approach for dealing with the divergences present when considering radiative corrections. 

Schwinger, Weisskopf and Oppenheimer suggested that the energy level shift might be the result of the interaction of the electron with the radiation field. The shift came out to be infinite in all theories at the time and was therefore ignored \cite{45}. 

Bethe \cite{45} provided the first proposal for a solution to the problem. His calculation diverged logarithmically so he introduced a cutoff $K$ for the light quanta which could be emitted and absorbed by the atom. He took $K$ to be $mc^2$ (where $m$ is the electron mass) and obtained a result of 1040 MHz which was within quite good agreement with the experimental value of $\approx1000$ MHz at the time \cite{45,51}. 

There followed rapidly a great deal of research by the pioneers of QFT such as Feynman \cite{50}, Schwinger \cite{61} and Dyson \cite{46} resulting in the renormalization program whereby the divergences in QFT can be circumvented.

The Lamb shift has continued to play a very important role in physics, both from the experimental point of view where its value has been determined to greater  and greater degrees of accuracy, and the theoretical point of view, where the theoretical prediction of its value has been refined to higher and higher orders of precision. The current level of agreement  between theory and experiment is as follows \cite{39}.
\begin{align}
\mbox{Experiment}:&1057.847(9)\\
\mbox{Theory$\,\,\,\,\,\,\,\,\,\,\,\,$}:&1057.834 12 (23)(13)
\end{align}
From the theoretical point of view research has developed on two fronts \cite{45,44,52,47,48,53,54}. Firstly approaches to formulating the problems and then for dealing with the troublesome divergences have been developed. Secondly numerical methods for computing the results of these theoretical approaches have been developed. The current state of the art description of the theory of the Lamb shift is given by Yerokhin \textit{et al.} in \cite{39}.

Research has also been carried out on hydrogenic ions and results are often presented as power series in $Z\alpha$. Self-energy for low $Z$ values is often obtained by extrapolating from high $Z$ calculations \cite{52}.

The leading order contributions to the Lamb shift are the self-energy contribution (electron self-energy) and the Uehling contribution (photon self-energy or vacuum polarization). The theoretical problem associated with the Uehling contribution to the Lamb shift was solved in 1935 by Uehling \cite{62} who derived the Uehling potential function. However his derivation using QFT as formulated at the time was, and all subsequent derivations using more modern techniques involving charge and mass  renormalization \cite{40} have been, complex and to a certain extent obscure during the process of negotiating divergences. Work on the Uehling contribution has focussed on developing numerical techniques for applying the Uehling potential function in the case of the point nucleus or more elaborate nuclear models. Theoretical work on the self-energy has taken various directions but there does not seem to have been developed an analogue for the self-energy to the Uehling potential function.

We have, in previous papers \cite{36,37} presented an approach called spectral regularization for analyzing problematic objects in QFT such as the vacuum polarization tensor $\Pi^{\mu\nu}$ or the self-energy function $\Sigma$ in terms of their existence as measures on Minkowski space. In the present paper we  provide some theoretical results on Lorentz invariant 2-tensor valued measures in Sections 2 and 3 and then in Section 4 derive the Uehling potential function without requiring renormalization. Then in Sections 5 and 6 we present some theoretical results concerning invariant ${\bf C}^{4\times4}$ valued measures and then in section 7 we derive a complex matrix valued potential function,  which is analogous to the Uehling potential function, and can be used to compute the self-energy contribution to the Lamb shift.

\section{Lorentz invariant 2-tensor valued measures}

Let ${\mathcal B}({\bf R}^4)$ denote the Borel algebra for ${\bf R}^4$. A tensor valued measure $\mu:{\mathcal B}({\bf R}^4)\rightarrow{\bf C}^{4\times4}$ will be said to be Lorentz invariant if
\begin{equation}
\mu^{\alpha\beta}(\Lambda\Upsilon)={\Lambda^{\alpha}}_{\rho}{\Lambda^{\beta}}_{\sigma}\mu^{\rho\sigma}(\Upsilon),\forall\Lambda\in O(1,3),\Upsilon\in{\mathcal B}({\bf R}^4).
\end{equation}

\subsection{Existence of a spectral function when the measure can be defined by a density with respect to Lebesgue measure}

Suppose that $\Pi^{\mu\nu}$ is a Lorentz invariant 2-tensor valued measure which is associated with a density which we also denote as $\Pi^{\mu\nu}$. Then
\begin{equation}
\Pi^{\mu\nu}(\Upsilon)=\int_{\Upsilon}\Pi^{\mu\nu}(p)\,dp,\forall\Upsilon\in{\mathcal B}({\bf R}^4).
\end{equation}
Thus, for any $\Lambda\in O(1,3)$,
\[ \Pi^{\mu\nu}(\Lambda\Upsilon)=\int_{\Lambda\Upsilon}\Pi^{\mu\nu}(p)\,dp=\int_{\Upsilon}\Pi^{\mu\nu}(\Lambda p)\,dp, \]
and
\[ \Pi^{\mu\nu}(\Lambda\Upsilon)={\Lambda^{\mu}}_{\rho}{\Lambda^{\nu}}_{\sigma}\Pi^{\rho\sigma}(\Upsilon)={\Lambda^{\mu}}_{\rho}{\Lambda^{\nu}}_{\sigma}\int_{\Upsilon}\Pi^{\rho\sigma}(p)\,dp, \]
for all $\Upsilon\in{\mathcal B}({\bf R}^4)$. Therefore for each $\Lambda\in O(1,3)$,
\begin{equation} \label{eq:density_1}
\Pi^{\mu\nu}(\Lambda p)={\Lambda^{\mu}}_{\rho}{\Lambda^{\nu}}_{\sigma}\Pi^{\rho\sigma}(p),
\end{equation}
for almost all $p\in{\bf R}^4$. We will consider the (non-pathological) case where  $\Pi$ can and has been adjusted so that Eq.~\ref{eq:density_1} holds for all $\Lambda\in O(1,3)$ and $p\in{\bf R}^4$.

Suppose also that $\Pi$ is causal and is supported on $\{p\in{\bf R}^4:p^2>0,p^0>0\}$. (The cone $\{p\in{\bf R}^4:p^2=0,p^0\ge0\}$ is a set of Lebesgue measure zero). Then $\Pi$ is determined by its values on $\{(m,{\vct 0}):m>0\}$. Define 
\begin{equation}
\lambda^{\mu\nu}(m)=\Pi^{\mu\nu}((m,{\vct 0})), \mbox{ for }m>0.
\end{equation}
Then
\begin{align*}
\lambda^{\mu\nu}(m)&=\Pi^{\mu\nu}((m,{\vct0}))\\
&=\Pi^{\mu\nu}(R(m,{\vct0}))\\
&={R^{\mu}}_{\rho}{R^{\nu}}_{\sigma}\Pi^{\rho\sigma}((m,{\vct0}))\\
&={R^{\mu}}_{\rho}{R^{\nu}}_{\sigma}\lambda^{\rho\sigma}(m),
\end{align*}
for all $m>0,R\in\mbox{Rotations}$, where Rotations $\cong O(3)$ is the rotation subgroup of $O(1,3)$. Then (suppressing the argument $m$)
\[ \lambda^{0\nu}={R^0}_{\rho}{R^{\nu}}_{\sigma}\lambda^{\rho\sigma}={R^{\nu}}_{\sigma}\lambda^{0\sigma},\forall R\in\mbox{ Rotations}. \]
It follows that 
\[ \lambda^{0\nu}=\lambda^{00}\eta^{0\nu}. \]
Similarly
\[ \lambda^{\mu0}=\lambda^{00}\eta^{\mu0}. \]
Therefore
\[ \lambda = \left(\begin{array}{cc}
\lambda^{00} & 0\\
0&A
\end{array}\right), \]
for some $A\in{\bf C}^{3\times3}$. Thus, since
\[ \lambda^{\mu\nu}={R^{\mu}}_{\rho}{R^{\nu}}_{\sigma}\lambda^{\rho\sigma},\forall R\in\mbox{ Rotations}, \]
we must have
\[ {\lambda^{ij}=B^{i}}_k{B^j}_l\lambda^{kl},\forall B\in O(3)\mbox{ where }i,j,k,l\in\{1,2,3\}. \]
Taking
\[ B=\left(\begin{array}{ccc}
0&1&0\\
-1&0&0\\
0&0&1
\end{array}\right), \]
shows that 
\[ \lambda^{12}={B^1}_{k}{B^{2}}_l\lambda^{kl}=-\lambda^{21}. \]
Taking
\begin{equation} \label{eq:density_matrix}
B=\left(\begin{array}{ccc}
0&1&0\\
1&0&0\\
0&0&1
\end{array}\right),
\end{equation}
shows that 
\[ \lambda^{12}={B^1}_{k}{B^{2}}_l\lambda^{kl}=\lambda^{21}. \]
Therefore
\[ \lambda^{12}=\lambda^{21}=0. \]
By a similar argument
\[ \lambda^{31}=\lambda^{13}=\lambda^{32}=\lambda^{23}=0. \]
Therefore for all $m>0$
\[ (\lambda^{\mu\nu}(m))=\mbox{diag}(\lambda_0(m),\lambda_1(m),\lambda_2(m),\lambda_3(m))\mbox{ for some }\lambda_0(m),\lambda_1(m),\lambda_2(m),\lambda_3(m)\in{\bf C}. \]
Now 
\[ \lambda^{11}={B^1}_k{B^1}_l\lambda^{kl}. \]
Taking $B$ to be of the form of Eq.~\ref{eq:density_matrix} results in
\[ \lambda^{11}={B^1}_k{B^1}_l\lambda^{kl}=\lambda^{22}. \]
Similarly
\[ \lambda^{11}=\lambda^{33}. \]
Thus
\begin{equation} \label{eq:density_general_form}
\lambda(m)=(\lambda^{\mu\nu}(m))=\mbox{diag}(\lambda_0(m),\lambda_1(m),\lambda_1(m),\lambda_1(m)),
\end{equation}
for some locally integrable functions $\lambda_0,\lambda_1:(0,\infty)\rightarrow{\bf C}$.

\subsection{Canonical form of a causal Lorentz invariant 2-tensor valued measure}

Define
\begin{equation}
\Pi_c^{\mu\nu}(\Upsilon)=\int_{m=0}^{\infty}\int_{\Upsilon}(\eta^{\mu\nu}p^2\sigma_1(m)+p^{\mu}p^{\nu}\sigma_2(m))\,\Omega_m(dp)\,dm,
\end{equation}
for $\Upsilon\in{\mathcal B}({\bf R}^4)$, where $\sigma_1,\sigma_2:(0,\infty)\rightarrow{\bf C}$ are locally integrable and for $m>0$ $\Omega_m$ is the standard Lorentz invariant measure on the hyperboloid $H_m=\{p\in{\bf R}^4:p^2=m^2,p^0>0\}$ defined by
\begin{equation}
\int\psi(p)\,\Omega_m(dp)=\int_{{\bf R}^3}\psi(\omega_m({\vct p}),{\vct p})\frac{d{\vct p}}{\omega_m({\vct p})},
\end{equation}
for $\psi$ a measurable function for which the integral exists, in which
\begin{equation}
\omega_m({\vct p})=(m^2+{\vct p}^2)^{\frac{1}{2}}.
\end{equation} 
Then
\begin{align*}
\Pi_c^{\mu\nu}(\Upsilon)&=\int_{m=0}^{\infty}\int_{{\vct p}\in{\bf R}^3}\chi_{\Upsilon}(\omega_m({\vct p}),{\vct p})(\eta^{\mu\nu}m^2\sigma_1(m)+(\omega_m({\vct p}),{\vct p})^{\mu}(\omega_m({\vct p}),{\vct p})^{\nu}\sigma_2(m))\\
&\frac{d{\vct p}}{\omega_m({\vct p})}\,dm,
\end{align*}
where $\chi_{\Gamma}$ denotes the characteristic function of $\Gamma$ defined by
\begin{equation}
\chi_{\Gamma}(p)=\left\{
\begin{array}{l}
1\mbox{ if }p\in\Gamma\\
0\mbox{ otherwise.}
\end{array}\right.
\end{equation}
Now make the coordinate transformation
\begin{equation} \label{eq:cood_xn}
q = q(m,{\vct p})=(\omega_m({\vct p}),{\vct p}),m>0,{\vct p}\in{\bf R}^3.
\end{equation}
The Jacobian of the transformation is 
\[ J(m,{\vct p})=m\omega_m({\vct p})^{-1}. \]
Therefore
\begin{align*}
\Pi_c^{\mu\nu}(\Upsilon)&=\int_{q^2>0,q^0>0}\chi_{\Upsilon}(q)(\eta^{\mu\nu}m^2\sigma_1(m)+q^{\mu}q^{\nu}\sigma_2(m))\frac{1}{\omega_m({\vct p})}\frac{1}{m\omega_m({\vct p})^{-1}}\,dq\\
&=\int_{q^2>0,q^0>0}\chi_{\Upsilon}(q)(\eta^{\mu\nu}m\sigma_1(m)+q^{\mu}q^{\nu}m^{-1}\sigma_2(m))\,dq,
\end{align*}
where $m=m(q)=(q^2)^{\frac{1}{2}}$. Therefore the density associated with the canonical form is
\begin{equation}
\Pi_c^{\mu\nu}(q)=\left\{\begin{array}{l}
\eta^{\mu\nu}\zeta(q)\sigma_1(\zeta(q))+q^{\mu}q^{\nu}\zeta(q)^{-1}\sigma_2(\zeta(q)), \mbox{ for } q^2>0,q^0>0\\
0\mbox{ otherwise,}
\end{array}\right.
\end{equation}
where $\zeta(q)=(q^2)^{\frac{1}{2}}$. Therefore the $\lambda$ function associated with $\Pi_c^{\mu\nu}$ is  
\begin{align*}
\lambda_c^{\mu\nu}(m)&=\Pi_c^{\mu\nu}((m,{\vct 0}))\\
&=\eta^{\mu\nu}m\sigma_1(m)+(m,{\vct 0})^{\mu}(m,{\vct 0})^{\nu}m^{-1}\sigma_2(m)\\
&=\eta^{\mu\nu}m\sigma_1(m)+\eta^{0\mu}\eta^{0\nu}m\sigma_2(m).
\end{align*}
Since $\sigma_1$ and $\sigma_2$ are arbitrary locally integrable functions, comparing with Eq.~\ref{eq:density_general_form} we see that this is in the most general form of a density for a causal Lorentz invariant 2-tensor valued measure. Hence the canonical form is the most general form.

\subsection{The spectral calculus for causal Lorentz invariant 2-tensor valued measures}

Let $\Pi^{\mu\nu}:{\mathcal B}({\bf R}^4)\rightarrow{\bf C}$ be a causal Lorentz invariant 2-tensor valued measure which can be defined by a locally integrable density. Then from the previous section, $\Pi^{\mu\nu}$ has the form
\begin{equation}
\Pi^{\mu\nu}(\Upsilon)=\int_{m=0}^{\infty}\int_{{\bf R}^4}\chi_{\Upsilon}(p)(p^2\eta^{\mu\nu}\sigma_1(m)+p^{\mu}p^{\nu}\sigma_2(m))\,\Omega_m(dp)\,dm, \label{eq:Pi_mu_nu_form}
\end{equation}
for some locally integrable spectral functions $\sigma_1,\sigma_2:(0,\infty)\rightarrow{\bf C}$. Assume that the spectral functions are continuous. Now let 
\begin{equation} \label{eq:Upsilon_a_b_def}
\Upsilon(a,b,\epsilon)=\bigcup_{m\in(a,b)}S(m,\epsilon),
\end{equation}
where
\begin{equation}
S(m,\epsilon)=\{p\in{\bf R}^4:p^2=m^2,|{\vct p}|<\epsilon,p^0>0\},
\end{equation}
be the hyperbolic cylinder defined in \cite{36}. 
Let 
\begin{equation}
g^{\mu\nu}(a,b,\epsilon)=\Pi^{\mu\nu}(\Upsilon(a,b,\epsilon)), \label{eq:g_mu_nu_def}
\end{equation}
for $a,b\in(0,\infty),a<b,\epsilon>0$. We have
\begin{align*}
g^{\mu\nu}(a,b,\epsilon)&=\Pi^{\mu\nu}(\Upsilon(a,b,\epsilon))\\
&=\int_{m=0}^{\infty}\int_{{\vct p}\in{\bf R}^3}\chi_{\Upsilon(a,b,\epsilon)}(\omega_m({\vct p}),{\vct p})[\eta^{\mu\nu}m^2\sigma_1(m)+(\omega_m({\vct p}),{\vct p})^{\mu}(\omega_m({\vct p}),{\vct p})^{\nu}\sigma_2(m)]\\
&\frac{d{\vct p}}{\omega_m({\vct p})}\,dm\\
&=\int_{m=a}^b\int_{B_{\epsilon}({\vct 0})}[\eta^{\mu\nu}m^2\sigma_1(m)+(\omega_m({\vct p}),{\vct p}))^{\mu}(\omega_m({\vct p}),{\vct p})^{\nu}\sigma_2(m)]\frac{d{\vct p}}{\omega_m({\vct p})}\,dm\\
&\approx\int_a^b[\eta^{\mu\nu}m^2\sigma_1(m)+\eta^{\mu0}\eta^{\nu0}m^2\sigma_2(m)]\frac{1}{m}\,dm\,(\frac{4}{3}\pi\epsilon^3).
\end{align*}
Therefore taking
\begin{align} \label{eq:g_mu_nu}
g^{\mu\nu}_a(b)&=\lim_{\epsilon\rightarrow0}\epsilon^{-3}g^{\mu\nu}(a,b,\epsilon),
\end{align}
we see that
\begin{align}
g_a^{\mu\nu}(b)&=\int_a^b[\eta^{\mu\nu}m\sigma_1(m)+\eta^{\mu0}\eta^{\nu0}m\sigma_2(m)]\,dm\,(\frac{4}{3}\pi).
\end{align}
Therefore we can recover the spectral functions $\sigma_1$ and $\sigma_2$ as follows.
\[ g_a^{ii\prime}(b)=-b\sigma_1(b)(\frac{4}{3}\pi),\forall i=1,2,3. \]
Therefore
\begin{equation}
\sigma_1(b)=-\frac{3}{4\pi}\frac{1}{b}g_a^{ii\prime}(b). \label{eq:density_5}
\end{equation}
Also
\begin{equation} \label{eq:density_4}
g_a^{00\prime}(b)=b(\sigma_1(b)+\sigma_2(b))(\frac{4}{3}\pi).
\end{equation}
Therefore
\begin{equation}
\sigma_2(b)=\frac{3}{4\pi}\frac{1}{b}g_a^{00\prime}(b)-\sigma_1(b). \label{eq:density_6}
\end{equation}
Conversely, if a causal Lorentz invariant measure $\Pi^{\mu\nu}$ is such that the functions $g_a^{\mu\nu}$ defined by Eq.~\ref{eq:g_mu_nu} exist and are continuously differentiable then $\Pi^{\mu\nu}$ has the form of Eq.~\ref{eq:Pi_mu_nu_form} and the spectral functions $\sigma_1$ and $\sigma_2$ can be recovered using Eqns.~\ref{eq:density_5}-\ref{eq:density_6}.

\section{The vacuum polarization tensor}

The vacuum polarization tensor is given by
\begin{align*}
\Pi^{\mu\nu}(k)&=-\mbox{Tr}(\int\frac{dp}{(2\pi)^{4}}i(-e)\gamma^{\mu}iS(p)i(-e)\gamma^{\nu}iS(p-k))\\
&=-\frac{e^2}{(2\pi)^{4}}\int\mbox{Tr}(\gamma^{\mu}\frac{1}{{\slas p}-m+i\epsilon}\gamma^{\nu}\frac{1}{{\slas p}-{\slas k}-m+i\epsilon})\,dp.
\end{align*}
(the leading minus sign is associated with the fermion loop). Therefore, using the {\em ansatz}  
\begin{equation*}
\frac{1}{p^2-m^2+i\epsilon}\rightarrow -i\pi\Omega_m(p),
\end{equation*}
and arguing as in \cite{36} the measure associated with the vacuum polarization tensor is given by 
\begin{equation}
\Pi^{\mu\nu}(\Upsilon)=-\frac{e^2}{16\pi^2}\int\chi_{\Upsilon}(k+p)\mbox{Tr}(\gamma^{\mu}({\slas p}+m)\gamma^{\nu}({\slas k}-m))\,\Omega_m(dk)\,\Omega_m(dp),
\end{equation}
for $\Upsilon\in{\mathcal B}({\bf R}^4)$.
\newtheorem{theorem}{Theorem}
\begin{theorem}
The vacuum polarization tensor valued measure is Lorentz invariant.
\end{theorem}
{\bf Proof} Rescale by $-\frac{e^2}{16\pi^2}$. Then
\begin{align*}
\Pi_{\mu\nu}(\Lambda(\kappa)\Upsilon)&=\int\chi_{\Lambda\Upsilon}(k+p)\mbox{Tr}(\gamma_{\mu}({\slas p}+m)\gamma_{\nu}({\slas k}-m))\,\Omega_m(dk)\,\Omega_m(dp) \\
&=\int\chi_{\Upsilon}(\Lambda^{-1}k+\Lambda^{-1}p)\mbox{Tr}(\gamma_{\mu}({\slas p}+m)\gamma_{\nu}({\slas k}-m))\,\Omega_m(dk)\,\Omega_m(dp) \\
&=\int\chi_{\Upsilon}(k+p)\mbox{Tr}(\gamma_{\mu}(\kappa{\slas p}\kappa^{-1}+m)\gamma_{\nu}(\kappa{\slas k}\kappa^{-1}-m))\,\Omega_m(dk)\,\Omega_m(dp) \\
&=\int\chi_{\Upsilon}(k+p)\mbox{Tr}(\kappa\kappa^{-1}\gamma_{\mu}\kappa({\slas p}+m)\kappa^{-1}\gamma_{\nu}\kappa({\slas k}-m)\kappa^{-1})\,\Omega_m(dk)\,\Omega_m(dp) \\
&=\int\chi_{\Upsilon}(k+p)\mbox{Tr}({\Lambda^{-1\rho}}_{\mu}\gamma_{\rho}({\slas p}+m){\Lambda^{-1\sigma}}_{\nu}\gamma_{\sigma}({\slas k}-m))\,\Omega_m(dk)\,\Omega_m(dp) \\
&={\Lambda^{-1\rho}}_{\mu}{\Lambda^{-1\sigma}}_{\nu}\int\chi_{\Upsilon}(k+p)\mbox{Tr}(\gamma_{\rho}({\slas p}+m)\gamma_{\sigma}({\slas k}-m))\,\Omega_m(dk)\,\Omega_m(dp) \\
&={\Lambda^{-1\rho}}_{\mu}{\Lambda^{-1\sigma}}_{\nu}\Pi_{\rho\sigma}(\Upsilon),
\end{align*}
for all $\kappa\in K$ where 
\begin{equation} \label{eq:K_def}
K=\{\left(
\begin{array}{cc}
a&0\\
0&a^{\dagger-1}
\end{array}\right):a\in GL(2,{\bf C}),|\mbox{det}(a)|=1\}\subset U(2,2),
\end{equation}
is the group defined in \cite{5}, $\Lambda=\Lambda(\kappa)$ is the Lorentz transformation corresponding to $\kappa$ and we have used the intertwining property $\Sigma_1(\kappa p)=\kappa\Sigma_1(p)\kappa^{-1}=\kappa{\slas p}\kappa^{-1}$ of the map $\Sigma_1=(p\mapsto{\slas p})$ \cite{5} which implies that $\kappa\gamma_{\mu}\kappa^{-1}=\Lambda(\kappa)^{\rho}{}_{\mu}\gamma_{\rho},\forall\kappa\in K$.

Therefore
\begin{align*}
\Pi^{\alpha\beta}(\Lambda\Upsilon)&=\eta^{\alpha\mu}\eta^{\beta\nu}\Pi_{\mu\nu}(\Lambda\Upsilon)\\
&=\eta^{\alpha\mu}\eta^{\beta\nu}{\Lambda^{-1\rho}}_{\mu}{\Lambda^{-1\sigma}}_{\nu}\Pi_{\rho\sigma}(\Upsilon).
\end{align*}
Now
\begin{equation}
{\Lambda^{-1\rho}}_{\mu}\eta^{\alpha\mu}=(\Lambda^{-1}\eta)^{\rho\alpha}, \nonumber
\end{equation}
and
\[ \Lambda^{T}\eta\Lambda=\eta, \]
from which it follows that
\[ \Lambda^{-1}\eta=\eta\Lambda^{T}. \]
Therefore
\[ {\Lambda^{-1\rho}}_{\mu}\eta^{\alpha\mu}=(\eta\Lambda^{T})^{\rho\alpha}=\eta^{\rho\mu}{\Lambda^{T}_{\mu}}^{\alpha}=\eta^{\rho\mu}{\Lambda^{\alpha}}_{\mu}. \]
Similarly
\[ {\Lambda^{-1\sigma}}_{\nu}\eta^{\beta\nu}=\eta^{\sigma\nu}{\Lambda^{\beta}}_{\nu}. \]
Therefore
\begin{align*}
\Pi^{\alpha\beta}(\Lambda\Upsilon)&=\eta^{\rho\mu}{\Lambda^{\alpha}}_{\mu}\eta^{\sigma\nu}{\Lambda^{\beta}}_{\nu}\Pi_{\rho\sigma}(\Upsilon)\\
&={\Lambda^{\alpha}}_{\mu}{\Lambda^{\beta}}_{\nu}\Pi^{\mu\nu}(\Upsilon).
\end{align*}
$\Box$

\subsection{Determination of the form of the vacuum polarization tensor using spectral calculus}

It is straightforward to show that the vacuum polarization tensor valued measure is causal. 
\begin{theorem}
The vacuum polarization tensor has the form
\[ \Pi^{\mu\nu}(\Upsilon)=\int_{m^{\prime}=0}^{\infty}\int_{{\bf R}^4}\chi_{\Upsilon}(p)(p^2\eta^{\mu\nu}-p^{\mu}p^{\nu})\,\Omega_{m^{\prime}}(dp)\,\rho(m^{\prime})\,dm^{\prime}, \]
for some continuous spectral function $\rho:(0,\infty)\rightarrow{\bf C}$.
\end{theorem}
{\bf Proof}

One can readily compute using the gamma matrix trace identities that
\[ \mbox{Tr}(\gamma^{\mu}({\slas p}+m)\gamma^{\nu}({\slas k}-m))=4(p^{\mu}k^{\nu}-\eta^{\mu\nu}p.k+k^{\mu}p^{\nu}-m^2\eta^{\mu\nu}). \]
Thus
\begin{align*}
g^{\mu\nu}(a,b,\epsilon)&=\Pi^{\mu\nu}(\Upsilon(a,b,\epsilon))\\
&=-\frac{e^2}{16\pi^2}\int\chi_{\Upsilon(a,b,\epsilon)}(k+p)\mbox{Tr}(\gamma^{\mu}({\slas p}+m)\gamma^{\nu}({\slas k}-m))\,\Omega_m(dk)\,\Omega_m(dp)\\
&=-\frac{e^2}{4\pi^2}\int\chi_{\Upsilon(a,b,\epsilon)}(k+p)(p^{\mu}k^{\nu}-\eta^{\mu\nu}p.k+k^{\mu}p^{\nu}-m^2\eta^{\mu\nu} )\,\Omega_m(dk)\,\Omega_m(dp)\\
&=-\frac{e^2}{4\pi^2}\int\chi_{(a,b)}(\omega_m({\vct k})+\omega_m({\vct p}))\chi_{B_{\epsilon}({\vct 0})}({\vct k}+{\vct p})((\omega_m({\vct p}),{\vct p})^{\mu}(\omega_m({\vct k}),{\vct k})^{\nu}-\\
&\eta^{\mu\nu}(\omega_m({\vct p})\omega_m({\vct k})-{\vct p}.{\vct k})+(\omega_m({\vct k}),{\vct k})^{\mu}(\omega_m({\vct p}),{\vct p})^{\nu}-m^2\eta^{\mu\nu})\,\frac{d{\vct k}}{\omega_m({\vct k})}\frac{d{\vct p}}{\omega_m({\vct p})}\\
&=-\frac{e^2}{4\pi^2}\int\chi_{(a,b)}(\omega_m({\vct k})+\omega_m({\vct p}))\chi_{B_{\epsilon}({\vct 0})-{\vct p}}({\vct k})((\omega_m({\vct p}),{\vct p})^{\mu}(\omega_m({\vct k}),{\vct k})^{\nu}-\\
&\eta^{\mu\nu}(\omega_m({\vct p})\omega_m({\vct k})-{\vct p}.{\vct k})+(\omega_m({\vct k}),{\vct k})^{\mu}(\omega_m({\vct p}),{\vct p})^{\nu}-m^2\eta^{\mu\nu})\,\frac{d{\vct k}}{\omega_m({\vct k})}\frac{d{\vct p}}{\omega_m({\vct p})}\\
&\approx-\frac{e^2}{4\pi^2}\int\chi_{(a,b)}(2\omega_m({\vct p}))((\omega_m({\vct p}),{\vct p})^{\mu}(\omega_m({\vct p}),-{\vct p})^{\nu}-\eta^{\mu\nu}(\omega_m({\vct p})^2+{\vct p}^2)+\\
&(\omega_m({\vct p}),-{\vct p})^{\mu}(\omega_m({\vct p}),{\vct p})^{\nu}-m^2\eta^{\mu\nu})\frac{1}{\omega_m({\vct p})^2}\,d{\vct p}\,(\frac{4}{3}\pi\epsilon^3)
\end{align*}
Now
\begin{align*}
\chi_{(a,b)}(2\omega_m({\vct p}))=1&\Leftrightarrow a<2(m^2+{\vct p}^2)^{\frac{1}{2}}<b\\
&\Leftrightarrow\frac{a^2}{4}<m^2+{\vct p}^2<\frac{b^2}{4}\\
&\Leftrightarrow\frac{a^2}{4}-m^2<{\vct p}^2<\frac{b^2}{4}-m^2\\
&\Leftrightarrow mZ(a)<|{\vct p}|<mZ(b),
\end{align*}
where
\begin{equation} \label{eq:Z_def}
Z(b)=(\frac{b^2}{4m^2}-1)^{\frac{1}{2}}, b\ge2m.
\end{equation}
Therefore
\begin{align*}
g_a^{\mu\nu}(b)&=\lim_{\epsilon\rightarrow0}\epsilon^{-3}g^{\mu\nu}(a,b,\epsilon)\\
&=-\frac{e^2}{4\pi^2}\int_{r=mZ(a)}^{mZ(b)}\int_{\theta=0}^{\pi}\int_{\phi=0}^{2\pi}((\omega_m({\vct p}),{\vct p})^{\mu}(\omega_m({\vct p}),-{\vct p})^{\nu}-\eta^{\mu\nu}(m^2+2r^2)+\\
&(\omega_m({\vct p}),-{\vct p})^{\mu}(\omega_m({\vct p}),{\vct p})^{\nu}-m^2\eta^{\mu\nu})\frac{r^2}{m^2+r^2}\sin(\theta)\,d\phi\,d\theta\,dr\,(\frac{4}{3}\pi).
\end{align*}
where ${\vct p}={\vct p}(r,\theta,\phi)$ (spherical polar coordinates). 
Hence
\begin{align*}
g_a^{\mu\mu\prime}(b)&=-\frac{e^2}{4\pi^2}\int_{\theta=0}^{\pi}\int_{\phi=0}^{2\pi}((\omega_m({\vct p}),{\vct p})^{\mu}(\omega_m({\vct p}),-{\vct p})^{\mu}-\eta^{\mu\mu}(m^2+2r^2)+\\
&(\omega_m({\vct p}),-{\vct p})^{\mu}(\omega_m({\vct p}),{\vct p})^{\mu}-m^2\eta^{\mu\mu})\frac{r^2}{m^2+r^2}\sin(\theta)\,d\phi\,d\theta\,(\frac{4}{3}\pi)\\
&mZ^{\prime}(b),
\end{align*}
where $r=mZ(b)$. Thus
\begin{align*}
g_a^{00\prime}(b)&=-\frac{e^2}{4\pi^2}\int_{\theta=0}^{\pi}\int_{\phi=0}^{2\pi}(m^2+r^2-(m^2+2r^2)+m^2+r^2-m^2)\\
&\frac{r^2}{m^2+r^2}\sin(\theta)\,d\phi\,d\theta\,(\frac{4}{3}\pi)(mZ^{\prime}(b))\\
&=0.
\end{align*}
Therefore by the spectral theorem and Eq.~\ref{eq:density_4},
$\Pi^{\mu\nu}$ has the form of Eq.~\ref{eq:Pi_mu_nu_form} with $\sigma_1$ and $\sigma_2$ continuous functions and 
\begin{equation}
\sigma_1(b)=-\sigma_2(b),\forall b>0.
\end{equation}
$\Box$

Thus, letting $\rho=\sigma_1$ and making the coordinate transformation defined by Eq.~\ref{eq:cood_xn}, we have that
\begin{align*}
\Pi^{\mu\nu}(\Upsilon)&=\int_{m^{\prime}=0}^{\infty}\int_{{\bf R}^3}\chi_{\Upsilon}(\omega_{m^{\prime}}({\vct p}),{\vct p})(\eta^{\mu\nu}m^{\prime2}-(\omega_{m^{\prime}}({\vct p}),{\vct p})^{\mu}(\omega_{m^{\prime}}({\vct p}),{\vct p})^{\nu})\rho(m^{\prime})\\
&\frac{d{\vct p}}{\omega_{m^{\prime}}({\vct p})}\,dm^{\prime}\\
&=\int_{q^2>0,q^0>0}\chi_{\Upsilon}(q)(q^2\eta^{\mu\nu}-q^{\mu}q^{\nu})(q^2)^{-\frac{1}{2}}\rho((q^2)^{\frac{1}{2}})\,dq.
\end{align*}
Hence the density $\Pi^{\mu\nu}$ for the measure $\Pi^{\mu\nu}$ is given by
\begin{equation}
\Pi^{\mu\nu}(q)=(q^2\eta^{\mu\nu}-q^{\mu}q^{\nu})\pi(q), \label{eq:Pi_form}
\end{equation}
where
\begin{equation}
\pi(q)=\left\{
\begin{array}{l}
(q^2)^{-\frac{1}{2}}\rho((q^2)^{\frac{1}{2}})\mbox{ for }q^2>0,q^0>0\\
0\mbox{ otherwise.}
\end{array}\right.
\end{equation}
(The fact that $\Pi^{\mu\nu}$ has the form of Eq.~\ref{eq:Pi_form} is well known but has previously been established through manipulating infinite quantities during renormalization (see \cite{70}, p. 478).) 

Contracting Eq.~\ref{eq:Pi_form} with the Minkowski space metric tensor we obtain
\begin{equation}
\pi(q)=\frac{1}{3q^2}\Pi(q)\mbox{ for }q^2\neq0.
\end{equation}
where
\begin{equation}
\Pi(q)=\eta_{\mu\nu}\Pi^{\mu\nu}(q).
\end{equation}

\subsection{Computation of the vacuum polarization function using spectral calculus}

It can be shown using the spectral calculus \cite{36,37}, that the contraction of the spectral vacuum polarization tensor $\Pi^{\mu\nu}$ is given by
\begin{equation}
\Pi(q)=\left\{\begin{array}{l}
s^{-1}\sigma(s)\mbox{ if }q^2>0, q^0>0\\
0\mbox{ otherwise,}
\end{array}\right.
\end{equation}
where $s=(q^2)^{\frac{1}{2}}$ and $\sigma$ is given by
\begin{equation}
\sigma(s)=\left\{
\begin{array}{l}
\frac{2}{\pi}e^2m^3Z(s)(3+2Z^2(s))\mbox{ if }s\geq2m\\
0\mbox{ otherwise,}
\end{array}\right.
\end{equation}
in which $Z$ is given by Eq.~\ref{eq:Z_def}.

Therefore the spectral vacuum polarization function is given by
\begin{equation}
\pi(q)=\frac{\Pi(q)}{3q^2}=\frac{1}{3}s^{-3}\sigma(s)=\frac{2}{3\pi}e^2m^3s^{-3}Z(s)(3+2Z^2(s)),\mbox{ if }s\ge2m,
\end{equation}
for $q^2>0,q^0>0$ where $s=(q^2)^{\frac{1}{2}}$.

It can also be shown by making the transformation $\Omega_m*\Omega_m\rightarrow\Omega_{im}*\Omega_{im}$ that, in the spacelike domain, $\pi$ can be considered to be the function given by
\begin{equation}
\pi(q)=\frac{1}{3}s^{-3}\sigma(s)=\frac{2}{3\pi}e^2m^3s^{-3}Z(s)(3+2Z^2(s)),\mbox{ if }s\ge2m,
\end{equation}
where $s=(-q^2)^{\frac{1}{2}}$.

\section{The Uehling contribution to the Lamb shift \label{section:Uehling_contribution}}

The Feynman amplitude associated with the tree diagram for electron proton scattering is given by
\begin{align*}
&{\mathcal M}_{a\alpha_1^{\prime}\alpha_2^{\prime}\alpha_1\alpha_2}(p_1^{\prime},p_2^{\prime},p_1,p_2)=\overline{u}_e(p_1^{\prime},\alpha_1^{\prime})ie_1\gamma^{\rho}u_e(p_1,\alpha_1)iD_{\rho\sigma}(q)\overline{u}_p(p_2^{\prime},\alpha_2^{\prime})ie_2\gamma^{\sigma}u_p(p_2,\alpha_2),
\end{align*}
where
\[ D_{F\alpha\beta}(q)=-\frac{\eta_{\alpha\beta}}{q^2+i\epsilon}, \]
is the photon propagator, $q=p_1^{\prime}-p_1$ is the momentum transfer, $e_1=-e,e_2=e$ are the charges of the electron and proton respectively and $u_e(p,\alpha)$ and $u_p(p,\alpha)$ are Dirac spinors for the electron and the proton for $\alpha\in\{0,1\}$. Therefore
\begin{equation}
{\mathcal M}_{a\alpha_1^{\prime}\alpha_2^{\prime}\alpha_1\alpha_2}(p_1^{\prime},p_2^{\prime},p_1,p_2)=i\frac{{\mathcal M}_{0\alpha_1^{\prime}\alpha_2^{\prime}\alpha_1\alpha_2}(p_1^{\prime},p_2^{\prime},p_1,p_2)}{q^2+i\epsilon},
\end{equation}
where
\begin{align*}
&{\mathcal M}_{0\alpha_1^{\prime}\alpha_2^{\prime}\alpha_1\alpha_2}(p_1^{\prime},p_2^{\prime},p_1,p_2)=\overline{u}_e(p_1^{\prime},\alpha_1^{\prime})e_1\gamma^{\rho}u_e(p_1,\alpha_1)\eta_{\rho\sigma}\overline{u}_p(p_2^{\prime},\alpha_2^{\prime})e_2\gamma^{\sigma}u_p(p_2,\alpha_2),
\end{align*}
The Feynman amplitude associated with the vacuum polarization diagram is given by
\begin{align*}
{\mathcal M}_{b\alpha_1^{\prime}\alpha_2^{\prime}\alpha_1\alpha_2}(p_1^{\prime},p_2^{\prime},p_1,p_2)&=\overline{u}_e(p_1^{\prime},\alpha_1^{\prime})ie_1\gamma^{\rho}u_e(p_1,\alpha_1)iD_{F\rho\mu}(q)i\Pi^{\mu\nu}(q)iD_{F\nu\sigma}(q)\\
&\overline{u}_p(p_2^{\prime},\alpha_2^{\prime})ie_2\gamma^{\sigma}u_p(p_2,\alpha_2),
\end{align*}
 Therefore, since
\begin{align*}
D_{F\rho\mu}(q)\Pi^{\mu\nu}(q)D_{F\nu\sigma}(q)&=(\frac{1}{q^2})^2\eta_{\rho\mu}(q^2\eta^{\mu\nu}-q^{\mu}q^{\nu})\pi(q)\eta_{\nu\sigma}, \\
&=(\frac{\eta_{\rho\sigma}}{q^2}-(\frac{1}{q^2})^2q_{\rho}q_{\sigma})\pi(q),
\end{align*}
and, by a well known conservation property
\[ \overline{u}(p_1^{\prime},\alpha_1^{\prime})q_{\rho}\gamma^{\rho}u(p_1,\alpha_1)=0, \]
we have that
\begin{equation} \nonumber
{\mathcal M}_{b\alpha_1^{\prime}\alpha_1^{\prime}\alpha_1\alpha_2}(p_1^{\prime},p_2^{\prime},p_1,p_2)=i\frac{{\mathcal M}_{0\alpha_1^{\prime}\alpha_1^{\prime}\alpha_1\alpha_2}(p_1^{\prime},p_2^{\prime},p_1,p_2)\pi(q)}{q^2+i\epsilon}.
\end{equation}
Now consider the NRQED (non-relativistic quantum electrodynamics) approximation in which
\begin{align*}
{\mathcal M}_{0\alpha_1^{\prime}\alpha_2^{\prime}\alpha_1\alpha_2}(p_1^{\prime},p_2^{\prime},p_1,p_2)&=e_1e_2\delta_{\alpha_1^{\prime}\alpha_1}\delta_{\alpha_2^{\prime},\alpha_2}=-e^2\delta_{\alpha_1^{\prime}\alpha_1}\delta_{\alpha_2^{\prime},\alpha_2},
\end{align*}
and $q^0$ is negligible compared with $|{\vct q}|$. In this case
\[ {\mathcal M}_{b\alpha_1^{\prime}\alpha_2^{\prime}\alpha_1\alpha_2}(p_1^{\prime},p_2^{\prime},p_1,p_2)=-ie^2\delta_{\alpha_1^{\prime}\alpha_1}\delta_{\alpha_2^{\prime}\alpha_2}\frac{\pi((0,{\vct q}))}{(0,{\vct q})^2}=ie^2\delta_{\alpha_1^{\prime}\alpha_1}\delta_{\alpha_2^{\prime}\alpha_2}\frac{\pi((0,{\vct q}))}{{\vct q}^2}, \]
and we can write
\begin{equation}
{\mathcal M}_b({\vct q})=ie^2\frac{\pi(0,{\vct q})}{{\vct q}^2}.
\end{equation}
Therefore using the Born approximation the potential function associated with the Uehling contribution to the Lamb shift is \cite{36}
\begin{align*}
\Delta V({\vct x})&=i(2\pi)^{-3}\int{\mathcal M}_b({\vct q})e^{i{\vct q}.{\vct x}}\,d{\vct q}\\
&=i(2\pi)^{-3}\int(ie^2\frac{\pi((0,{\vct q}))}{{\vct q}^2})e^{i{\vct q}.{\vct x}}\,d{\vct q}\\
&=-(2\pi)^{-3}e^2\int\frac{\pi((0,{\vct q}))}{{\vct q}^2}e^{i{\vct q}.{\vct x}}\,d{\vct q},
\end{align*}
if the integral exists. The argument of the inverse Fourier transform is orthogonally invariant and thus $\Delta V$ is orthogonally invariant. Therefore we can write
\begin{align*}
\Delta V(r)&=\Delta V(0,0,r)\\
&=-(2\pi)^{-3}e^2\int_{s=0}^{\infty}\int_{\theta=0}^{\pi}\int_{\phi=0}^{2\pi}\frac{\pi(s)}{s^2}e^{isr\cos(\theta)}s^2\sin(\theta)\,d\phi\,d\theta\,ds\\
&=-(2\pi)^{-2}e^2\int_{s=0}^{\infty}\frac{\pi(s)}{irs}(e^{isr}-e^{-isr})\,ds\\
&=i(2\pi)^{-2}e^2\frac{1}{r}\int_{s=0}^{\infty}\frac{\pi(s)}{s}(e^{isr}-e^{-isr})\,ds.
\end{align*}
Hence
\begin{equation} \label{eq:Delta_V}
\Delta V(r)=i(2\pi)^{-2}e^2\frac{1}{r}(\int_{s=0}^{\infty}\frac{\pi(s)}{s}e^{isr}\,ds-\int_{s=0}^{\infty}\frac{\pi(s)}{s}e^{-isr}\,ds).
\end{equation}
It is shown in \cite{36} that these integrals are convergent for all $r>0$. We would like to analytically continue $\Delta V$ to the upper imaginary axis of the complex plane since we interested in spacelike points in Minkowski (configuration) space for which $x^2<0$, corresponding to pure imaginary $r$. Therefore we seek a complex analytic function $\Delta V_{\mbox{analytic}}$ associated with $\Delta V$. 
By a well known Paley-Weiner theorem the function
\[ r\mapsto\int_{s=0}^{\infty}\frac{\pi(s)}{rs}e^{irs}\,ds, \]
is analytic in the upper half plane while the function
\[ r\mapsto\int_{s=0}^{\infty}\frac{\pi(s)}{rs}e^{-irs}, \]
is analytic in the lower half plane (but not in the upper half plane). 

Consider the transformation ${\mathcal T}$ taking functions $F$ of the form 
\begin{equation} \label{eq:transform_1}
F(r)=\int_{s=0}^{\infty}f(s)(e^{isr}-e^{-isr})\,ds,
\end{equation}
to the function ${\mathcal T}(F)$ defined by
\begin{equation} \label{eq:transform_2}
({\mathcal T}(F))(r)=i\int_{s=0}^{\infty}f(s)e^{-isr}\,ds.
\end{equation}
where $f:(0,\infty)\rightarrow{\bf C}$ is such that both integrals defined by Eqns.~\ref{eq:transform_1} and \ref{eq:transform_2} are convergent for all $r>0$.
${\mathcal T}$ takes non-divergent functions into non-divergent functions. The mathematical properties and justification of the transform ${\mathcal T}$ may be considered elsewhere. The important thing from the point of view of physics is that the use of ${\mathcal T}$ gives the correct answers. We find that it does for the cases of the computation of the Uehling potential function and also for the case of the electron self-energy contribution to the Lamb shift.

In the case of the Ueling contribution we compute using 
\begin{equation}
f(s)=\frac{\pi(s)}{s}=\frac{\sigma(s)}{3s^4},
\end{equation}
and the $\Delta V$ function that we derive is
\begin{align*}
\Delta V(r)&=i(2\pi)^{-2}e^2\frac{1}{r}F(r)\\
&\rightarrow i(2\pi)^{-2}e^2\frac{1}{r}({\mathcal T}(F))(r)\\
&=i(2\pi)^{-2}e^2\frac{1}{r}i\int_{s=0}^{\infty}\frac{\sigma(s)}{3s^4}e^{-sr}\,ds\\
&=-(2\pi)^{-2}e^2\frac{2}{3\pi}e^2m^3\frac{1}{r}\int_{s=2m}^{\infty}\frac{Z(s)(3+2Z^2(s))}{s^4}e^{-sr}\,ds\\
&=-(2\pi)^{-2}e^2\frac{2}{3\pi}e^2m^3\frac{1}{r}\int_{x=1}^{\infty}\frac{(x^2-1)^{\frac{1}{2}}(2x^2+1)}{(16m^4)x^4}e^{-2mxr}(2m)\,dx\\
&=-\frac{\alpha^2}{3\pi}\frac{1}{r}\int_{x=1}^{\infty}(x^2-1)^{\frac{1}{2}}(2x^2+1)x^{-4}e^{-2mxr}\,dx,
\end{align*}
where $\alpha=\frac{e^2}{4\pi}$ is the fine structure constant. 

The result that we have obtained is precisely the Uehling vacuum polarization potential function which has previously only been calculated through negotiating infinities and divergences using complex calculations involving methods such as charge and mass renormalization \cite{40}. 

From this potential function the Uehling contribution to the Lamb shift can be exactly calculated \cite{38} according to
\begin{equation}
\Delta E=<\psi|\Delta V|\psi>=4\pi\int_{r=0}^{\infty}|\psi(r)|^2\Delta V(r)r^2\,dr,
\end{equation}
where $\psi$ is the H atom 2s wave function, and theory agrees with experiment to a very high order of precision \cite{39}.  

\section{$K$ invariant ${\bf C}^{4\times4}$ matrix valued measures}

As described in \cite{5} the group $K\subset U(2,2)\subset{\bf C}^{4\times4}$ defined by Eq.~\ref{eq:K_def} acts in natural ways on ${\bf R}^4$ and ${\bf C}^4$. A ${\bf C}^{4\times4}$ matrix valued measure is a vector valued measure taking values in the vector space ${\bf C}^{4\times4}$. A ${\bf C}^{4\times4}$ valued measure $\mu:{\mathcal B}({\bf  R}^4)\rightarrow{\bf C}^{4\times4}$ will be said to be $K$ invariant if
\begin{equation}
\mu(\kappa\Upsilon)=\kappa\mu(\Upsilon)\kappa^{-1},\forall\kappa\in K,\Upsilon\in{\mathcal B}({\bf R}^4).
\end{equation}

\subsection{Existence of a spectral function when the measure can be defined by a density with respect to Lebesgue measure \label{section:general_K_invt_form}}
 
Suppose that $\mu$ is such a measure which can be defined by a locally integrable density, which we also denote as $\mu$. Then
\begin{equation} \label{eq:matrix_density}
\mu(\Upsilon)=\int_{\Upsilon}\mu(p)\,dp,\forall\Upsilon\in{\mathcal B}({\bf R}^4).
\end{equation}
By $K$ invariance
\begin{equation}
\mu(\Lambda(\kappa)\Upsilon)=\kappa\mu(\Upsilon)\kappa^{-1}=\kappa\int_{\Upsilon}\mu(p)\,dp\,\kappa^{-1}.
\end{equation}
But
\begin{equation}
\mu(\Lambda(\kappa)\Upsilon)=\int_{\Lambda(\kappa)\Upsilon}\mu(p)\,dp=\int_{\Upsilon}\mu(\Lambda(\kappa)p)\,dp,
\end{equation}
where $\Lambda(\kappa)$ is the element of $O(1,3)^{\uparrow+}$ corresponding to $\kappa\in K$. Both these equations are true for all $\kappa\in K,\Upsilon\in{\mathcal B}({\bf R}^4)$. Therefore for all $\kappa\in K$
\begin{equation} \label{eq:condition10}
\mu(\Lambda(\kappa)p)=\kappa\mu(p)\kappa^{-1},
\end{equation}
(for almost all $p\in{\bf R}^4$). We will consider the case where $\mu$ can be, and has been, adjusted so that Eq.~\ref{eq:condition10} holds for all $\kappa\in K,p\in{\bf R}^4$.

Conversely, given a locally integrable matrix valued function which satisfies Eq.~\ref{eq:condition10} then the object $\mu:{\mathcal B}({\bf R}^4)\rightarrow{\bf C}^{4\times4}$ defined by Eq.~\ref{eq:matrix_density} is a $K$ invariant ${\bf C}^{4\times4}$ valued measure. 

Suppose that $\mu$ is such a function and assume that $\mu$ is causal and is supported on $\{p\in{\bf R}^4:p^2>0, p^0>0\}$.  Define $M:(0,\infty)\rightarrow{\bf C}^{4\times4}$ by
\begin{equation}
M(m)=\mu((m,{\vct 0})). \label{eq:M_def}
\end{equation}
Now
\begin{equation}
\mu(\Lambda(\kappa)(m,{\vct 0}))=\kappa M(m)\kappa^{-1}.
\end{equation}
Thus $\mu$ is determined if $M$ is given. We will call the function $M$ the spectrum of $\mu$.

Since $(m,{\vct 0})$ is invariant under elements of Rotations we have
\begin{eqnarray}
M(m) & = & \left(\begin{array}{ll}
M_1 & M_2 \\
M_3 & M_4
\end{array}\right) \nonumber \\
    & = & \left(\begin{array}{ll}
a & 0 \\
0 & a
\end{array}\right)\left(\begin{array}{ll}
M_1 & M_2 \\
M_3 & M_4
\end{array}\right)\left(\begin{array}{ll}
a & 0 \\
0 & a
\end{array}\right)^{-1} \nonumber \\
& = & \left(\begin{array}{ll}
aM_1a^{-1} & aM_2a^{-1} \\
aM_3a^{-1} & aM_4a^{-1}
\end{array}\right), \nonumber
\end{eqnarray}
for all $a\in SU(2)$. Therefore each block $M_i,i=1,2,3,4$ of $M$ commutes with each element of $SU(2)$. Suppose that 
\begin{equation}
M_i=\left(\begin{array}{cc}
b_{i11} & b_{i12} \\
b_{i21} & b_{i22}
\end{array}\right).
\end{equation}
Now
\begin{equation}
a=\left(\begin{array}{cc}
0 & 1 \\
-1 & 0
\end{array}\right)\in SU(2).
\end{equation}
Therefore
\begin{equation}
\left(\begin{array}{cc}
0 & 1 \\
-1 & 0
\end{array}\right)
\left(\begin{array}{cc}
b_{i11} & b_{i12} \\
b_{i21} & b_{i22}
\end{array}\right)=
\left(\begin{array}{cc}
b_{i11} & b_{i12} \\
b_{i21} & b_{i22}
\end{array}\right)
\left(\begin{array}{cc}
0 & 1 \\
-1 & 0
\end{array}\right),
\end{equation}
from which it follows that $b_{i21}=-b_{i12}$ and $b_{i11}=b_{i22}$. Also
\begin{equation}
a=\left(\begin{array}{cc}
0 & i \\
i & 0
\end{array}\right)\in SU(2).
\end{equation}
Therefore
\begin{equation}
\left(\begin{array}{cc}
0 & i \\
i & 0
\end{array}\right)
\left(\begin{array}{cc}
b_{i11} & b_{i12} \\
b_{i21} & b_{i22}
\end{array}\right)=
\left(\begin{array}{cc}
b_{i11} & b_{i12} \\
b_{i21} & b_{i22}
\end{array}\right)
\left(\begin{array}{cc}
0 & i \\
i & 0
\end{array}\right),
\end{equation}
from which it follows that $b_{i21}=b_{i12}$ and $b_{i11}=b_{i22}$. Therefore
\begin{equation}
M_i=\lambda_i=\lambda_iI_2,
\end{equation}
for some $\lambda_i\in{\bf C}$.

Conversely, let $\lambda_1,\lambda_2,\lambda_3,\lambda_4:(0,\infty)\rightarrow{\bf C}$ be locally integrable functions and define $\mu:\{p\in{\bf R}^4:p^2>0,p^0>0\}\rightarrow{\bf C^{4\times4}}$ by
\begin{equation} \label{eq:general_form_of_K_invt_mu}
\mu(\kappa(m,{\vct 0}))=\kappa M(m)\kappa^{-1},\forall\kappa\in K,m>0,
\end{equation}
where
\begin{equation}
M=\left(
\begin{array}{cc}
\lambda_1&\lambda_2\\
\lambda_3&\lambda_4
\end{array}\right)\in{\bf C}^{4\times4}.
\end{equation}
It is straightforward to show that $\mu$ is well defined. 
Let $p\mapsto\kappa(p)$ be any function such that $\kappa(p)p=((p^2)^{\frac{1}{2}},{\vct 0}),\forall p\in{\bf R}^4\mbox{ with }p^2>0,p^0>0$. Then for all $\kappa\in K,p\in{\bf R}^4$ for which $p^2>0,p^0>0$ 
\begin{align*}
\mu(\kappa p)&=\mu(\kappa\kappa(p)^{-1}\kappa(p) p)\\
&=\mu(\kappa\kappa(p)^{-1}(m,{\vct 0}))\\
&=\kappa\kappa(p)^{-1}M(m)(\kappa\kappa(p)^{-1})^{-1}\\
&=\kappa\mu(\kappa(p)^{-1}(m,{\vct0}))\kappa^{-1}\\
&=\kappa\mu(\kappa(p)^{-1}\kappa(p) p)\kappa^{-1}\\
&=\kappa\mu(p)\kappa^{-1}.
\end{align*}
where $m=(p^2)^{\frac{1}{2}}$. Therefore $\mu$ is $K$ invariant.

Hence the function defined by Eq.~\ref{eq:general_form_of_K_invt_mu} is the most general form of the density for a $K$ invariant ${\bf C}^{4\times4}$ valued measure which can be defined by a locally integrable density on Minkowski space.

\subsection{Canonical form of a $K$ invariant ${\bf C}^{4\times4}$ valued measure}

If $\sigma:(0,\infty)\rightarrow{\bf C}$ is a locally integrable function define $\mu_{\sigma}:{\mathcal B}({\bf R}^4)\rightarrow{\bf C}^{4\times4}$ by
\begin{equation}
\mu_{\sigma}(\Upsilon)=\int_{m=0}^{\infty}\int_{\Upsilon}({\slas p}+m)\,\Omega_m(dp)\sigma(m)\,dm.
\end{equation}
Then
\begin{eqnarray}
\mu_{\sigma}(\Lambda(\kappa)(\Upsilon)) & = & \int_{m=0}^{\infty}\int_{\Lambda(\kappa)(\Upsilon)}({\slas p}+m)\,\Omega_m(dp)\sigma(m)\,dm \nonumber \\
     & = & \int_{m=0}^{\infty}\int_{\Upsilon}(\Sigma_1(\Lambda(\kappa)p)+m)\,\Omega_m(dp)\sigma(m)\,dm \nonumber \\
     & = & \int_{m=0}^{\infty}\int_{\Upsilon}(\kappa{\slas p}\kappa^{-1}+m)\,\Omega_m(dp)\sigma(m)\,dm \nonumber \\
     & = & \kappa\int_{m=0}^{\infty}\int_{\Upsilon}({\slas p}+m)\,\Omega_m(dp)\sigma(m)\,dm\,\kappa^{-1} \nonumber \\
    & = & \kappa\mu(\Upsilon)\kappa^{-1}, \nonumber
\end{eqnarray}
for all $\Upsilon\in{\mathcal B}({\bf R}^4)$ where $\Sigma_1$ denotes the map $p\mapsto{\slas p}$ and we have used the intertwining property $\Sigma_1(\kappa p)=\kappa\Sigma_1(p)\kappa^{-1}$ of $\Sigma_1$ \cite{5}. Therefore $\mu_{\sigma}$ is $K$ invariant. Now, making the coordinate transformation defined by Eq.~\ref{eq:cood_xn}, we have 
\begin{eqnarray}
\mu_{\sigma}(\Upsilon) & = & \int_{m=0}^{\infty}\int_{\Upsilon}({\slas p}+m)\,\Omega_m(dp)\sigma(m)\,dm \nonumber \\
    & = & \int_{m=0}^{\infty}\int_{{\bf R}^3}\chi_{\Upsilon}((\omega_m({\vct p}),{\vct p}))(\Sigma_1((\omega_m({\vct p}),{\vct p}))+m)\frac{d{\vct p}}{\omega_m({\vct p})}\sigma(m)\,dm \nonumber \\
    & = & \int_{q^2>0,q^0>0}\chi_{\Upsilon}(q)({\slas q}+\zeta(q))\frac{\sigma(\zeta(q))}{\zeta(q)}\,dq, \nonumber
\end{eqnarray}
where $\zeta(q)=(q^2)^{\frac{1}{2}}$. Therefore the density associated with $\mu_{\sigma}$ is given by
\begin{equation}
\mu_{\sigma}(q)=\left\{\begin{array}{l}
({\slas q}+\zeta(q))\zeta(q)^{-1}\sigma(\zeta(q))\mbox{ if }q^2>0,q^0>0 \\
0\mbox{ otherwise.}
\end{array}\right.
\end{equation}
Therefore the spectral function $M=M_{\sigma}:(0,\infty)\rightarrow{\bf C}^{4\times4}$ associated with $\mu_{\sigma}$ is
\begin{eqnarray}
M_{\sigma}(m) & = & \mu_{\sigma}((m,{\vct 0})) \nonumber \\
    & = & (m\gamma^0+m)m^{-1}\sigma(m)\nonumber\\
    & =  & (\gamma^0+1)\sigma(m)\nonumber\\
    & = & \left(\begin{array}{cc}
2\sigma(m) & 0 \\
0 & 0
\end{array}\right),\nonumber
\end{eqnarray}
 where we use the Dirac representation for the gamma matrices.

More generally if $\sigma_1,\sigma_2:(0,\infty)\rightarrow{\bf C}$ are locally integrable functions then the matrix valued measure $\mu_{\sigma_1,\sigma_2}:{\mathcal B}({\bf R}^4)\rightarrow{\bf C}^{4\times4}$ defined by
\begin{equation} \label{eq:general_K_invt_form}
\mu_{\sigma_1,\sigma_2}(\Upsilon)=\int_{m=0}^{\infty}\int_{\Upsilon}({\slas p}+m)\,\Omega_m(dp)\sigma_1(m)\,dm+\int_{m=0}^{\infty}\int_{\Upsilon}({\slas p}-m)\,\Omega_m(dp)\sigma_2(m)\,dm, 
\end{equation}
is $K$ invariant with spectral function $M$ given by
\begin{equation}
M_{\sigma_1,\sigma_2}(m)=\left(\begin{array}{cc}
2\sigma_1(m) & 0 \\
0 & -2\sigma_2(m)
\end{array}\right).
\end{equation}

\subsection{The spectral calculus for $K$ invariant ${\bf C}^{4\times4}$ valued measures}

Suppose that $\mu$ is a $K$ invariant ${\bf C}^{4\times4}$ valued measure which can be defined by a locally integrable density of the form of $\mu_{\sigma_1\sigma_2}$ for some locally integrable spectral functions $\sigma_1,\sigma_2$. Suppose that $\sigma_1$ and $\sigma_2$ are continuous on $(0,\infty)$. Let for $a,b,\epsilon\in(0,\infty),a<b$, $g(a,b,\epsilon)$ be defined by
\begin{equation}
g(a,b,\epsilon)=\mu(\Upsilon(a,b,\epsilon)),
\end{equation}
where $\Upsilon(a,b,\epsilon)$ is the hypercylinder defined by Eq.~\ref{eq:Upsilon_a_b_def}. 
Then
\begin{eqnarray}
g(a,b,\epsilon) & = & \int_{m=0}^{\infty}\int_{\Upsilon(a,b,\epsilon)}({\slas p}+m)\,\Omega_m(dp)\sigma_1(m)\,dm+ \nonumber \\
    &  & \int_{m=0}^{\infty}\int_{\Upsilon(a,b,\epsilon)}({\slas p}-m)\,\Omega_m(dp)\sigma_2(m)\,dm \nonumber \\
    & = & \int_{m=a}^b\int_{B_{\epsilon}({\vct 0})}(\Sigma_1((\omega_m({\vct p}),{\vct p}))+m)\,\frac{d{\vct p}}{\omega_m({\vct p}}\sigma_1(m)\,dm+ \nonumber \\
    &  & \int_{m=a}^b\int_{B_{\epsilon}({\vct 0})}(\Sigma_1((\omega_m({\vct p}),{\vct p}))-m)\,\frac{d{\vct p}}{\omega_m({\vct p})}\sigma_2(m)\,dm \nonumber \\
& \approx & \frac{4}{3}\pi\epsilon^3(\int_{m=a}^b(m\gamma^0+m)\,\frac{1}{m}\sigma_1(m)\,dm+ \nonumber \\
    &  & \int_{m=a}^b(m\gamma^0-m)\,\frac{1}{m}\sigma_2(m)\,dm) \nonumber \\
    & = & \frac{4}{3}\pi\epsilon^3\int_a^b\left(\begin{array}{cc}
\sigma_1(m) & 0 \\
0 & -\sigma_2(m)\end{array}\right)dm,
\end{eqnarray}
where $\Sigma_1$ denotes the map $p\mapsto{\slas p}$. Therefore if we define $g_a:(0,\infty)\rightarrow{\bf C}^{4\times4}$ by
\begin{equation} \label{eq:limit_g_b}
g_a(b)=\lim_{\epsilon\rightarrow0}\epsilon^{-3}g(a,b,\epsilon),
\end{equation}
then
\begin{equation} \label{eq:K_invt_spectra}
\sigma_1(b)=\frac{3}{4\pi}g_a^{00\prime}(b),\sigma_2(b)=-\frac{3}{4\pi}g_a^{33\prime}(b), \mbox{ for }b>0.
\end{equation}
i.e.
\begin{equation}
\sigma(b)=M(b)=\frac{3}{4\pi}g_a^{\prime}(b).
\end{equation}

Conversely if $\mu:{\mathcal B}({\bf R}^4)\rightarrow{\bf C}^{4\times4}$ is a causal $K$ invariant measure and if the function $g_a$ defined by Eq.~\ref{eq:limit_g_b} exists and is is continuously differentiable with $g_a^{\alpha\beta}(b)=0\mbox{ for }\alpha\neq\beta$  then $\mu$ has the form of Eq.~\ref{eq:general_K_invt_form} and the spectral functions can be recovered using Eqns.~\ref{eq:K_invt_spectra}.
 
\section{The self-energy of the electron}

The Feynman integral associated with the self-energy of the electron is
\begin{equation}
i\Sigma(p)=\int\frac{dk}{(2\pi)^4}iD_{\mu\nu}(k)i(-e)\gamma^{\mu}iS_F(p-k)i(-e)\gamma^{\nu},
\end{equation}
where
\begin{equation}
D_{\mu\nu}(k)=-\frac{1}{k^2+i\epsilon},
\end{equation}
is the photon propagator and
\begin{equation}
S_F(p)=\frac{1}{{\slas p}-m+i\epsilon},
\end{equation}
is the fermion propagator. This can be written as
\begin{equation}
i\Sigma(p)=-\frac{e^2}{(2\pi)^4}\int\frac{1}{k^2+i\epsilon}\gamma^{\mu}\frac{{\slas p}-{\slas k}+m}{(p-k)^2-m^2+i\epsilon}\gamma_{\mu}\,dk.
\end{equation}
We make the following formal computation
\begin{eqnarray}
i\Sigma(\Upsilon) & = & \int_{\Upsilon}i\Sigma(p)\,dp \nonumber \\
    & = & \int\chi_{\Upsilon}(p)(-\frac{e^2}{(2\pi)^4})\frac{1}{k^2+i\epsilon}\gamma^{\mu}\frac{{\slas p}-{\slas k}+m}{(p-k)^2-m^2+i\epsilon}\gamma_{\mu}\,dk\,dp \nonumber \\
    & = & \int\chi_{\Upsilon}(p)(-\frac{e^2}{(2\pi)^4})\frac{1}{k^2+i\epsilon}\gamma^{\mu}\frac{{\slas p}-{\slas k}+m}{(p-k)^2-m^2+i\epsilon}\gamma_{\mu}\,dp\,dk \nonumber \\
    & = & -\frac{e^2}{(2\pi)^4}\int\chi_{\Upsilon}(p+k)\frac{1}{k^2+i\epsilon}\gamma^{\mu}\frac{{\slas p}+m}{p^2-m^2+i\epsilon}\gamma_{\mu}\,dp\,dk \nonumber \\
    & = & \frac{e^2}{16\pi^2}\int\chi_{\Upsilon}(p+k)\gamma^{\mu}({\slas p}+m)\gamma_{\mu}\,\Omega_m^{\pm}(dp)\,\Omega_0^{\pm}(dk), \nonumber
\end{eqnarray}
where we have used the {\em ansatz} \cite{22}
\begin{equation}
\frac{1}{p^2-m^2+i\epsilon}\rightarrow-i\pi\Omega_m^{\pm}(p),\forall m\ge0.
\end{equation}
We take the case 
\begin{equation}
i\Sigma(\Upsilon)=\frac{e^2}{16\pi^2}\int\chi_{\Upsilon}(p+k)\gamma^{\mu}({\slas p}+m)\gamma_{\mu}\,\Omega_m(dp)\,\Omega_0^{+}(dk),
\end{equation}
in which case $\Sigma$ has existence as a well defined mathematical object (a tempered ${\bf C}^{4\times4}$ valued measure).

Now $\gamma^{\mu}{\slas p}\gamma_{\mu}=-2{\slas p}$ and $\gamma^{\mu}m\gamma_{\mu}=4m$. Therefore
\begin{equation}
i\Sigma(\Upsilon)=\frac{e^2}{16\pi^2}\int\chi_{\Upsilon}(p+k)(4m-2{\slas p})\,\Omega_m(dp)\,\Omega_0^{+}(dk).
\end{equation}
\begin{theorem}
$i\Sigma$ is $K$ invariant.
\end{theorem}
{\bf Proof}

Let $\kappa\in K$ and $\Upsilon\in{\mathcal B}({\bf R}^4)$. Then
\begin{eqnarray}
i\Sigma(\kappa(\Upsilon)) & = & \frac{e^2}{16\pi^2}\int\chi_{\kappa(\Upsilon)}(p+k)(4m-2{\slas p})\,\Omega_m(dp)\,\Omega_0^{+}(dk) \nonumber \\
    & = & \frac{e^2}{16\pi^2}\int\chi_{\Upsilon}(\kappa^{-1}(p+k))(4m-2{\slas p})\,\Omega_m(dp)\,\Omega_0^{+}(dk) \nonumber \\
    & = & \frac{e^2}{16\pi^2}\int\chi_{\Upsilon}(p+k)(4m-2{\kappa\slas p}\kappa^{-1})\,\Omega_m(dp)\,\Omega_0^{+}(dk) \nonumber \\
    & = & \frac{e^2}{16\pi^2}\kappa\int\chi_{\Upsilon}(p+k)(4m-2{\slas p})\,\Omega_m(dp)\,\Omega_0^{+}(dk)\kappa^{-1} \nonumber \\
    & = & \kappa i\Sigma(\Upsilon)\kappa^{-1} \nonumber
\end{eqnarray}
$\Box$

It can be shown that $i\Sigma$ is causal. We will now use the spectral calculus to compute the spectrum of $i\Sigma$.
\begin{eqnarray}
g(a,b,\epsilon)&=&i\Sigma(\Upsilon(a,b,\epsilon))\nonumber\\ 
& = & \frac{e^2}{16\pi^2}\int\chi_{\Upsilon(a,b,\epsilon)}(p+k)(4m-2{\slas p})\,\Omega_m(dp)\,\Omega_0^{+}(dk) \nonumber \\
    & = & \frac{e^2}{16\pi^2}\int\chi_{(a,b)}(\omega_m({\vct p})+\omega_0({\vct k}))\chi_{B_{\epsilon}({\vct 0})}({\vct p}+{\vct k})(4m-2{\slas p}) \nonumber \\
    &  & \omega_m({\vct p})^{-1}\omega_0({\vct k})^{-1}\,d{\vct p}\,d{\vct k} \nonumber \\
    & = & \frac{e^2}{16\pi^2}\int\chi_{(a,b)}(\omega_m({\vct p})+\omega_0({\vct k}))\chi_{B_{\epsilon}({\vct 0})-{\vct p}}({\vct k})(4m-2{\slas p}) \nonumber \\
    &  & \omega_m({\vct p})^{-1}\omega_0({\vct k})^{-1}\,d{\vct p}\,d{\vct k} \nonumber \\
   & = & \frac{e^2}{16\pi^2}\int\chi_{(a,b)}(\omega_m({\vct p})+\omega_0({\vct k}))\chi_{B_{\epsilon}({\vct 0})-{\vct p}}({\vct k})(4m-2{\slas p}) \nonumber \\
   &  & \omega_m({\vct p})^{-1}\omega_0({\vct k})^{-1}\,d{\vct k}\,d{\vct p} \nonumber \\
   & \approx & \frac{4}{3}\pi\epsilon^3\frac{e^2}{16\pi^2}\int\chi_{(a,b)}(\omega_m({\vct p})+\omega_0({\vct p}))(4m-2{\slas p}) \nonumber \\
    &  & \omega_m({\vct p})^{-1}\omega_0({\vct p})^{-1}\,d{\vct p}, \nonumber
\end{eqnarray}
with $p=(\omega_m({\vct p}),{\vct p})$.

Now $\omega_m({\vct p})+\omega_0({\vct p})=c\Rightarrow c\ge m$, and for all $c\ge m$
\begin{align*}
\omega_m({\vct p})+\omega_0({\vct p})=c&\Leftrightarrow(r^2+m^2)^{\frac{1}{2}}+r=c\\
&\Leftrightarrow r^2+m^2=(c-r)^2=c^2+r^2-2cr\\
&\Leftrightarrow2cr=c^2-m^2\\
&\Leftrightarrow r=(2c)^{-1}(c^2-m^2),
\end{align*}
where $r=|{\vct p}|$. Therefore
\begin{eqnarray}
g_a(b) & = & \lim_{\epsilon\rightarrow0}\epsilon^{-3}g(a,b,\epsilon)\nonumber\\
&=&\frac{4}{3}\pi\frac{e^2}{16\pi^2}\int\chi_{(a,b)}(\omega_m({\vct p})+\omega_0({\vct p}))(4m-2{\slas p})\omega_m({\vct p})^{-1}\omega_0({\vct p})^{-1}\,d{\vct p} \nonumber \\
    & = & \frac{4}{3}\pi\frac{e^2}{16\pi^2}\int_{r=Z_1(a)}^{Z_1(b)}\int_{\theta=0}^{\pi}\int_{\phi=0}^{2\pi}(4m-2{\slas p})\omega_m(r)^{-1}\omega_0(r)^{-1}r^2\sin(\theta)\,d\phi\,d\theta\,dr, \nonumber \\
\end{eqnarray}
where $p=p(r,\theta,\phi)=(\omega_m(r),r\sin\theta\cos(\phi),r\sin(\theta)\sin(\phi),r\cos(\theta))$ and $Z_1:[m,\infty)\rightarrow[0,\infty)$ is defined by
\begin{equation}
Z_1(b)=(2b)^{-1}(b^2-m^2).
\end{equation}
Now
\begin{equation}
Z_1^{\prime}(b)=(2b^2)^{-1}(b^2+m^2),
\end{equation}
and
\[ \int_{\theta=0}^{\pi}\int_{\phi=0}^{2\pi}{\slas p}\sin(\theta)\,d\phi\,d\theta=4\pi\omega_m(r)\gamma^0. \]
Hence
\begin{eqnarray}
g_a(b) & = & \frac{4}{3}\pi\frac{e^2}{16\pi^2}\int_{r=Z_1(a)}^{Z_1(b)}(4\pi)(4m-2\omega_m(r)\gamma^0)\omega_m(r)^{-1}r\,dr. \nonumber
\end{eqnarray}
Therefore applying the spectral calculus and using Leibniz' integral rule we obtain
\begin{align*}
\sigma(b) &=\frac{3}{4\pi}g_a^{\prime}(b)\\ 
&=\frac{e^2}{4\pi}(4m-2\omega_m(Z_1(b))\gamma^0)\omega_m(Z_1(b))^{-1}Z_1(b)Z_1^{\prime}(b)\mbox{ for }b\ge m.
\end{align*}
If $p\in{\bf R}^4$ is timelike and $\kappa\in K$ is such that $\kappa p=\Lambda(\kappa)p=((p^2)^{\frac{1}{2}},{\vct0})$ then 
\begin{equation}
i\Sigma(p)=i\Sigma(\kappa^{-1}\kappa p)=\kappa^{-1}i\Sigma(((p^2)^{\frac{1}{2}},{\vct0}))\kappa=\kappa^{-1}\sigma((p^2)^{\frac{1}{2}})\kappa.
\end{equation}
Also, in the spacelike domain, if $p\in{\bf R}^4$ is spacelike and $\kappa\in K$ is such that $\kappa p=(0,0,0,(-p^2)^{\frac{1}{2}})$ then
\begin{equation}
i\Sigma(p)=i\Sigma(\kappa^{-1}\kappa p)=\kappa^{-1}i\Sigma((0,0,0,(-p^2)^{\frac{1}{2}}))\kappa=\kappa^{-1}\sigma((-p^2)^{\frac{1}{2}})\kappa.
\end{equation}
In particular
\begin{equation}
i\Sigma((0,0,0,\zeta))=\sigma(\zeta),\forall\zeta>0.
\end{equation}

\section{The electron self-energy contribution to the Lamb shift}

The Feynman amplitude for the electron self-energy contribution to the Lamb shift is given by
\begin{equation}
{\mathcal M}={\mathcal M}_0+{\mathcal M}_1,
\end{equation}
where
\begin{eqnarray}
{\mathcal M}_{0\alpha_1^{\prime}\alpha_2^{\prime}\alpha_1\alpha_2}(p_1^{\prime},p_2^{\prime},p_1,p_2) & = & \overline{u}_e(p_2^{\prime},\alpha_2^{\prime})iD_{\mu\nu}(p_2^{\prime}-p_2)ie_2\gamma^{\mu}u_e(p_2,\alpha_2) \nonumber \\
    &   & \overline{u}_p(p_1^{\prime},\alpha_1^{\prime})ie_1\gamma^{\nu}u_p(p_1,\alpha_1), \nonumber \\
{\mathcal M}_{1\alpha_1^{\prime}\alpha_2^{\prime}\alpha_1\alpha_2}(p_1^{\prime},p_2^{\prime},p_1,p_2) & = & \overline{u}_e(p_2^{\prime},\alpha_2^{\prime})i\Sigma(p_2^{\prime})iS(p_2^{\prime})iD_{\mu\nu}(p_2^{\prime}-p_2)ie_2\gamma^{\mu}u_e(p_2,\alpha_2) \nonumber \\
    &   & \overline{u}_p(p_1^{\prime},\alpha_1^{\prime})ie_1\gamma^{\nu}u_p(p_1,\alpha_1), \nonumber 
\end{eqnarray}
in which $e_1=e$ and $e_2=-e$ are the charges of the proton and the electron respectively,
\begin{equation}
u_e(p,\alpha)=m_e^{-1}({\slas p}+m_e)e_{\alpha},
\end{equation}
\begin{equation}
u_p(p,\alpha)=m_p^{-1}({\slas p}+m_p)e_{\alpha},
\end{equation}
are Dirac spinors for the electron and the proton respectively with $m_e=$ the mass of the electron, $m_p=$ the mass of the proton, $e_{\alpha}$ is the $\alpha$th standard basis element for ${\bf C}^4$,
\begin{equation}
D_{\mu\nu}(q)=-\frac{\eta_{\mu\nu}}{q^2+i\epsilon},
\end{equation}
is the photon propagator,
\begin{equation}
S(p)=\frac{1}{{\slas p}-m_e+i\epsilon}=\frac{{\slas p}+m_e}{p^2-m_e^2+i\epsilon},
\end{equation}
is the electron propagator and
\begin{equation}
i\Sigma(p)``="-\frac{e^2}{(2\pi)^4}\int\frac{1}{k^2+i\epsilon}\gamma^{\mu}\frac{{\slas p}-{\slas k}+m_e}{(p-k)^2-m_e^2+i\epsilon}\gamma_{\mu}\,dk,
\end{equation}
is the the function associated with the Feynman integral for the self-energy of the electron.

${\mathcal M}_0$ corresponds to the tree level Feynman amplitude for electron-proton scattering and ${\mathcal M}_1$ is the perturbation due to the electron self-energy. Now
\begin{align*}
{\mathcal M}_{1\alpha_1^{\prime}\alpha_2^{\prime}\alpha_1\alpha_2}(p_1^{\prime},p_2^{\prime},p_1,p_2)&=\overline{u}_e(p_2^{\prime},\alpha_2^{\prime})i\Sigma(p_2^{\prime})S(p_2^{\prime})D_{\mu\nu}(p_2^{\prime}-p_2)e_2\gamma^{\mu}u_e(p_2,\alpha_2)\\
&\overline{u}_p(p_1^{\prime},\alpha_1^{\prime})e_1\gamma^{\nu}u_p(p_1,\alpha_1)\\
&=m_e^{-2}m_p^{-2}e_2e_1\overline{({\slas p_2^{\prime}}+m_e)e_{\alpha_2^{\prime}}}i\Sigma(p_2^{\prime})S(p_2^{\prime})D_{\mu\nu}(p_2^{\prime}-p_2)\gamma^{\mu}\\
&({\slas p_2}+m_e)e_{\alpha_2}\overline{({\slas p_1^{\prime}}+m_p)e_{\alpha_1^{\prime}}}\gamma^{\nu}({\slas p_1}+m_p)e_{\alpha_1}\\
&=m_e^{-2}m_p^{-2}e_2e_1e_{\alpha_2^{\prime}}^{\dagger}(\gamma^0{\slas p_2^{\prime}}\gamma^0+m_e)\gamma^0i\Sigma(p_2^{\prime})S(p_2^{\prime})D_{\mu\nu}(p_2^{\prime}-p_2)\gamma^{\mu}\\
&({\slas p_2}+m_e)e_{\alpha_2}e_{\alpha_1^{\prime}}^{\dagger}(\gamma^0{\slas p_1^{\prime}}\gamma^0+m_p)\gamma^0\gamma^{\nu}({\slas p_1}+m_{p})e_{\alpha_1}\\
&=m_e^{-2}m_p^{-2}e_2e_1e_{\alpha_2^{\prime}}^{\dagger}\gamma^0({\slas p_2^{\prime}}+m_e)i\Sigma(p_2^{\prime})S(p_2^{\prime})D_{\mu\nu}(p_2^{\prime}-p_2)\gamma^{\mu}\\
&({\slas p_2}+m_e)e_{\alpha_2}e_{\alpha_1^{\prime}}^{\dagger}\gamma^0({\slas p_1^{\prime}}+m_p)\gamma^{\nu}({\slas p_1}+m_{p})e_{\alpha_1}.
\end{align*}
Raising indices using the metric tensor $g=\gamma^0$ \cite{22} we obtain
\begin{align*}
{{{\mathcal M}_1}^{\alpha_1^{\prime}\alpha_2^{\prime}}}_{\alpha_1\alpha_2}(p_1^{\prime},p_2^{\prime},p_1,p_2)&=m_e^{-2}m_p^{-2}e_2e_1e_{\alpha_2^{\prime}}^{\dagger}({\slas p_2^{\prime}}+m_e)i\Sigma(p_2^{\prime})S(p_2^{\prime})D_{\mu\nu}(p_2^{\prime}-p_2)\gamma^{\mu}\\
&({\slas p_2}+m_e)e_{\alpha_2}e_{\alpha_1^{\prime}}^{\dagger}({\slas p_1^{\prime}}+m_p)\gamma^{\nu}({\slas p_1}+m_{p})e_{\alpha_1}.
\end{align*}
In \cite{22} the notion of a covariant kernel was defined. We have the following.
\begin{theorem}
${\mathcal M}_1$ is a covariant kernel.
\end{theorem}
{\bf Proof}
Let $\kappa\in K$. Then
\begin{align*}
{{{\mathcal M}_1}^{\alpha_1^{\prime}\alpha_2^{\prime}}}_{\alpha_1\alpha_2}(\kappa p_1^{\prime},\kappa p_2^{\prime},\kappa p_1,\kappa p_2)&=m_e^{-2}m_p^{-2}e_2e_1e_{\alpha_2^{\prime}}^{\dagger}({\kappa\slas p_2^{\prime}\kappa^{-1}}+m_e)\kappa i\Sigma(p_2^{\prime})\kappa^{-1}\kappa S(p_2^{\prime})\kappa^{-1}\\
&D_{\mu\nu}(\kappa p_2^{\prime}-\kappa p_2)\gamma^{\mu}(\kappa{\slas p_2}\kappa^{-1}+m_e)e_{\alpha_2}e_{\alpha_1^{\prime}}^{\dagger}(\kappa{\slas p_1^{\prime}}\kappa^{-1}+m_p)\gamma^{\nu}\\
&(\kappa{\slas p_1}\kappa^{-1}+m_{p})e_{\alpha_1}\\
&=-m_e^{-2}m_p^{-2}e_2e_1e_{\alpha_2^{\prime}}^{\dagger}\kappa({\slas p_2^{\prime}}+m_e)i\Sigma(p_2^{\prime})S(p_2^{\prime})\eta^{\mu\nu}q^{-2}\kappa^{-1}\gamma_{\mu}\\
&\kappa({\slas p_2}+m_e)\kappa^{-1}e_{\alpha_2}e_{\alpha_1}^{\dagger}\kappa({\slas p_1^{\prime}}+m_p)\kappa^{-1}\gamma_{\nu}\kappa({\slas p_1}+m_p)\kappa^{-1}e_{\alpha_1}\\
&=-m_e^{-2}m_p^{-2}e_2e_1e_{\alpha_2^{\prime}}^{\dagger}\kappa({\slas p_2^{\prime}}+m_e)i\Sigma(p_2^{\prime})S(p_2^{\prime})\eta^{\mu\nu}q^{-2}\Lambda^{-1\rho}{}_{\mu}\gamma_{\rho}\\
&({\slas p_2}+m_e)\kappa^{-1}e_{\alpha_2}e_{\alpha_1}^{\dagger}\kappa({\slas p_1^{\prime}}+m_p){\Lambda^{-1\sigma}}_{\nu}\gamma_{\sigma}({\slas p_1}+m_p)\kappa^{-1}e_{\alpha_1}\\
&=-m_e^{-2}m_p^{-2}e_2e_1e_{\alpha_2^{\prime}}^{\dagger}\kappa({\slas p_2^{\prime}}+m_e)i\Sigma(p_2^{\prime})S(p_2^{\prime})\eta^{\rho\sigma}\\
&q^{-2}\gamma_{\rho}({\slas p_2}+m_e)\kappa^{-1}e_{\alpha_2}e_{\alpha_1}^{\dagger}\kappa({\slas p_1^{\prime}}+m_p)\gamma_{\sigma}({\slas p_1}+m_p)\kappa^{-1}e_{\alpha_1}\\
&=-m_e^{-2}m_p^{-2}e_2e_1\\
&[\kappa({\slas p_2^{\prime}}+m_e)i\Sigma(p_2^{\prime})S(p_2^{\prime})\eta^{\rho\sigma}q^{-2}\gamma_{\rho}({\slas p_2}+m_e)\kappa^{-1}]^{\alpha_2^{\prime}}{}_{\alpha_2}\\
&{[\kappa({\slas p_1^{\prime}}+m_p)\gamma_{\sigma}({\slas p_1}+m_p)\kappa^{-1}]^{\alpha_1^{\prime}}}_{\alpha_1}\\
&=-m_e^{-2}m_p^{-2}e_2e_1\\
&{\kappa^{\alpha_2^{\prime}}}_{\beta_2^{\prime}}[({\slas p_2^{\prime}}+m_e)i\Sigma(p_2^{\prime})S(p_2^{\prime})\eta^{\rho\sigma}q^{-2}\gamma_{\rho}({\slas p_2}+m_e)]^{\beta_2^{\prime}}{}_{\beta_2}\kappa^{-1\beta_2}{}_{\alpha_2}\\
&\kappa^{\alpha_1^{\prime}}{}_{\beta_1^{\prime}}[({\slas p_1^{\prime}}+m_p)\gamma_{\sigma}({\slas p_1}+m_p)]^{\beta_1^{\prime}}{}_{\beta_1}\kappa^{-1\beta_1}{}_{\alpha_1}\\
&={\kappa^{\alpha_2^{\prime}}}_{\beta_2^{\prime}}\kappa^{-1\beta_2}{}_{\alpha_2}\kappa^{\alpha_1^{\prime}}{}_{\beta_1^{\prime}}\kappa^{-1\beta_1}{}_{\alpha_1}{{{\mathcal M}_1}^{\beta_1^{\prime}\beta_2^{\prime}}}_{\beta_1\beta_2}(p_1^{\prime},p_2^{\prime},p_1,p_2),
\end{align*}
where $\Lambda=\Lambda(\kappa)$ is the Lorentz transformation associated with $\kappa$ and $q=p_2^{\prime}-p_2$ is the momentum transfer for the scattered electron. Here we have used the intertwining property $\Sigma_1(\kappa p)=\kappa\Sigma_1(p)\kappa^{-1},\forall\kappa\in K,p\in{\bf R}^4$ where $\Sigma_1$ denotes the map $p\mapsto{\slas p}$ \cite{5}.
$\Box$

Our problem is considerably simplified if we make a non-relativistic approximation for the behaviour of the nucleus of the H atom since the proton is comparatively heavy and does not move much. In this approximation Dirac spinors $u_p(p,\alpha)$ satify
\begin{equation}
\overline{u}_p(p_1^{\prime},\alpha_1^{\prime})\gamma^{\mu}u_p(p_1,\alpha_1)=\delta_{\alpha_1^{\prime}\alpha_1}\eta^{\mu0},
\end{equation}
see \cite{36}. This implies that
\begin{equation}
m_p^{-2}({\slas p_1^{\prime}}+m_p)\gamma^{\mu}({\slas p_1}+m_p)=\eta^{\mu0}\gamma^0.
\end{equation}
Therefore the Feynman amplitude ${\mathcal M}_1$ for the electron self-energy contribution to electron-proton scattering (in the NR approximation for the proton) is
\begin{align*}
{\mathcal M_1}^{\alpha^{\prime}}{}_{\alpha}(p^{\prime},p) & = e^2m_e^{-2}e_{\alpha^{\prime}}^{\dagger}({\slas p^{\prime}}+m_e)i\Sigma(p^{\prime})S(p^{\prime})q^{-2}\gamma^{0}({\slas p}+m_e)e_{\alpha},
\end{align*}
where $p$ is the incoming electron momentum, $p^{\prime}$ is the outgoing electron momentum, $q=p^{\prime}-p$ is the momentum transfer and we have suppressed the proton polarization indices which play no further part in the calculation. Thus the Feynman amplitude matrix valued function is given by
\begin{equation}
{\mathcal M}_1={\mathcal M}_1(p^{\prime},p)=e^2m_e^{-2}({\slas p^{\prime}}+m_e)i\Sigma(p^{\prime})S(p^{\prime})q^{-2}\gamma^{0}({\slas p}+m_e).
\end{equation}
${\mathcal M}_1$ is not covariant with respect to all of the group $K$ which is not surprising since we are using an NR approximation. However we have the following .
\begin{theorem}
${\mathcal M}_1$ is covariant with respect to the rotation subgroup of $K$, that is, the group
\begin{equation}
\mbox{Rotations}=\{\left(
\begin{array}{cc}
a&0\\
0&a
\end{array}\right):a\in U(2)\}\subset K.
\end{equation}
\end{theorem}
{\bf Proof}

Let $R\in$ Rotations. Then $R^{-1}=R^{\dagger}$. Thus
\begin{align*}
{\mathcal M}_1(Rp^{\prime},Rp)&=e^2m_e^{-2}(R{\slas p^{\prime}}R^{-1}+m_e)Ri\Sigma(p^{\prime})R^{-1}RS(p^{\prime})R^{-1}(Rq)^{-2}\gamma^{0}(R{\slas p}R^{-1}+m_e)\\
&=Re^2m_e^{-2}({\slas p^{\prime}}+m_e)i\Sigma(p^{\prime})S(p^{\prime})q^{-2}R^{-1}\gamma^0R({\slas p}+m_e)R^{-1}\\
&=Re^2m_e^{-2}({\slas p^{\prime}}+m_e)i\Sigma(p^{\prime})S(p^{\prime})q^{-2}\gamma^0({\slas p}+m_e)R^{-1}\\
&=R{\mathcal M}_1(p^{\prime},p)R^{-1},
\end{align*}
 since
\[ R^{-1}\gamma^0R=R^{\dagger}\gamma^0R=\gamma^0. \]
($R\in K\subset U(2,2)$ and the metric for $U(2,2)$ is $g=\gamma^0$.)
We have used the intertwining property of the Feynman slash and the fact that the Feynman fermion propagator $S$ is $K$ invariant \cite{22}. 

$\Box$

Now carry out a translation in momentum space to the center of mass frame with origin $c=\frac{1}{2}(p+p^{\prime})$. Then 
\begin{align*}
p&\rightarrow p-c=\frac{1}{2}(p-p^{\prime})=-\frac{1}{2}q,\\
p^{\prime}&\rightarrow p^{\prime}-c=\frac{1}{2}(p^{\prime}-p)=\frac{1}{2}q.
\end{align*}
In this frame
\begin{equation}
{\mathcal M}_1(q)={\mathcal M}_1(p^{\prime},p)=q^{-2}{\mathcal N}(\frac{1}{2}q),
\end{equation}
where
\begin{equation}
{\mathcal N}(q)=e^2m_e^{-2}({\slas q}+m_e)i\Sigma(q)S(q)\gamma^{0}({-\slas q}+m_e).
\end{equation}
${\mathcal N}$ is rotationally covariant in the sense that
\begin{equation}
{\mathcal N}(Rq)=R{\mathcal N}(q)R^{-1},\forall R\in \mbox{Rotations}.
\end{equation}

Now $K$ acts on $C^{\infty}({\bf R}^4,{\bf C}^4)$ according to \cite{22}
\begin{equation}
(\kappa\psi)(x)=\kappa\psi(\kappa^{-1}x)=\kappa\psi(\Lambda(\kappa)^{-1}x),
\end{equation}
where $\Lambda(\kappa)$ is the Lorentz transformation corresponding to $\kappa$.
\newtheorem{lemma}{Lemma}
\begin{lemma} \label{lemma:Dirac_1}
Let $A\in C^{\infty}({\bf R}^4,{\bf C}^4)$ transform as a 4-vector under Lorentz transformations. Then
\begin{equation}
\kappa({\slas A}\psi)={\slas A}(\kappa\psi),\forall\kappa\in K,\psi\in C^{\infty}({\bf R}^4,{\bf C}^4).
\end{equation}
\end{lemma}
{\bf Proof}\newline
Let $x\in{\bf R}^4$. Then
\begin{align*}
(\kappa({\slas A}\psi))(x)&=\kappa({\slas A}\psi)(\Lambda^{-1}x)\\
&=\kappa{\slas A}(\Lambda^{-1}x)\psi(\Lambda^{-1}x)\\
&=\kappa\gamma^{\mu}A_{\mu}(\Lambda^{-1}x)\psi(\Lambda^{-1}x)\\
&=\kappa\gamma_{\mu}A^{\mu}(\Lambda^{-1}x)\psi(\Lambda^{-1}x)\\
&=\kappa\gamma_{\mu}{\Lambda^{-1\mu}}_{\nu}A^{\nu}(x)\psi(\Lambda^{-1}x)\\
&=\kappa\Sigma_1(\Lambda^{-1}A(x))\psi(\Lambda^{-1}x)\\
&=\kappa\kappa^{-1}{\slas A}(x)\kappa\psi(\Lambda^{-1}x)\\
&={\slas A}(x)(\kappa\psi)(x)\\
&=({\slas A}(\kappa\psi))(x).
\end{align*}
$\Box$\newline
Similarly
\begin{lemma} \label{lemma:Dirac_2}
\begin{equation}
\kappa({\slas\partial}\psi)={\slas\partial}(\kappa\psi),\forall\kappa\in K,\psi\in C^{\infty}({\bf R}^4,{\bf C}^4).
\end{equation}
\end{lemma}
\begin{theorem} \label{theorem:Dirac}
Let $\psi\in C^{\infty}({\bf R}^4,{\bf C}^4)$ be a unique solution to the Dirac equation with respect to a set ${\mathcal B}$ of boundary conditions. Then
\begin{equation}
\psi(\kappa x)=\kappa\psi(x),\forall\kappa\in K,x\in{\bf R}^4.
\end{equation}
\end{theorem}
{\bf Proof}

The Dirac equation is
\begin{equation}
(i\gamma^{\mu}D_{\mu}-m)\psi=0,
\end{equation}
where
\begin{equation}
D_{\mu}=\partial_{\mu}+ieA_{\mu}.
\end{equation}
Let $\psi\in C^{\infty}({\bf R}^4,{\bf C}^4)$ be the unique solution with respect to ${\mathcal B}$. Then
\begin{equation} \label{eq:Dirac_1}
(i{\slas D}-m)\psi=0.
\end{equation}
Therefore by Lemmas~\ref{lemma:Dirac_1} and~\ref{lemma:Dirac_2} 
\begin{equation}
0=\kappa(i{\slas D}-m)\psi=i\kappa({\slas\partial}\psi)-e\kappa({\slas A}\psi)-\kappa m\psi=i{\slas\partial}(\kappa\psi)-e{\slas A}(\kappa\psi)-m\kappa\psi=(i{\slas D}-m)(\kappa\psi). \nonumber
\end{equation}
But $\psi$ is uniquely determined by Eq.~\ref{eq:Dirac_1} and ${\mathcal B}$. Therefore ${\kappa\psi}=\psi$ and hence
\[ \kappa\psi(\Lambda(\kappa)^{-1}x)=\psi(x),\forall x\in{\bf R}^4, \]
from which it follows that
\begin{equation}
\psi(\Lambda(\kappa)x)=\kappa\psi(x),\forall\kappa\in K,x\in{\bf R}^4.
\end{equation}

(We have assumed that the transformed boundary conditions ${\ch{\mathcal B}}$ agree with ${\mathcal B}$.)
$\Box$

Now suppose that we have a static rotationally invariant Dirac eigenfunction $\psi$ for the 2s state of the H atom.
In general suppose that we have a Feynman amplitude ${\mathcal M}$ e.g. the tree level amplitude ${\mathcal M}_0$ or the amplitude ${\mathcal M}_0+{\mathcal M}_1$ where ${\mathcal M}_1$ is the electron self-energy amplitude. We propose that the energy associated with the bound state defined by $\psi$ is given by the integral
\begin{equation}
E=\omega\int\psi^{\dagger}({\vct x}){\mathcal M}({0,\vct q})\psi({\vct x})e^{i{\vct q}.{\vct x}}\,d{\vct x}\,d{\vct q},
\end{equation}
where we will determine $\omega$ by examination of the tree level diagram in the NR approximation. For this diagram 
\begin{equation}
{\mathcal M}_0(q)=ie^2|{\vct q}|^{-2}.
\end{equation}
However, as is well known, Feynman amplitudes are only defined up to multiplication by an element of $U(1)$ since physical predictions are made on the basis of the modulus squared of the Feynman amplitude. We choose to take
\begin{equation}
{\mathcal M}_0(q)=e^2|{\vct q}|^{-2},
\end{equation}
Then
\begin{align*}
E_0&=\omega\int\psi^{\dagger}({\vct x}){\mathcal M}_0(0,{\vct q})\psi({\vct x})e^{i{\vct q}.{\vct x}}\,d{\vct q}\,d{\vct x}\\
&=\omega e^2\int|\psi({\vct x})|^2(\int(|{\vct q}|^{-2}e^{i{\vct q}.{\vct x}}\,d{\vct q})\,d{\vct x}.
\end{align*}
But as is well known
\begin{equation}
\int|{\vct q}|^{-2}e^{i{\vct q}.{\vct x}}\,d{\vct q}=\frac{2\pi^2}{|{\vct x}|}.
\end{equation}
Therefore
\begin{equation}
E_0=8\pi^3\omega e^2\int_{r=0}^{\infty}|\psi(r)|^2r\,dr.
\end{equation}
However we know that
\begin{equation}
E_0=4\pi\int_{r=0}^{\infty}|\psi(r)|^2V_0(r)r^2\,dr=4\pi\int_{r=0}^{\infty}|\psi(r)|^2(-\frac{e^2}{4\pi r})r^2\,dr=-e^2\int_{r=0}^{\infty}|\psi(r)|^2r\,dr.
\end{equation}
Therefore $8\pi^3\omega=-1$ and thus 
\begin{equation}
\omega=-(2\pi)^{-3}.
\end{equation}
Now the perturbation of the H atom energy level due to the electron self energy is given by
\begin{equation}
\Delta E=\omega\int F({\vct q})\,d{\vct q},
\end{equation}
where
\begin{equation}
F({\vct q})=\int\psi^{\dagger}({\vct x}){\mathcal M}_1(0,{\vct q})\psi({\vct x})e^{i{\vct q}.{\vct x}}\,d{\vct x}.
\end{equation}
\begin{theorem} \label{theorem:rot_inv}
$F$ is rotationally invariant.
\end{theorem}
{\bf Proof}
\begin{align*}
F(A{\vct q})&=\int\psi^{\dagger}({\vct x}){\mathcal M}_1(0,A{\vct q})\psi({\vct x})e^{iA{\vct q}.{\vct x}}\,d{\vct x}\\
&=\int\psi^{\dagger}({\vct x})R{\mathcal M}_1(0,{\vct q})R^{\dagger}\psi({\vct x})e^{i{\vct q}.A^{\dagger}{\vct x}}\,d{\vct x}\\
&=\int\psi^{\dagger}({A^{\dagger}\vct x}){\mathcal M}_1(0,{\vct q})\psi(A^{\dagger}{\vct x})e^{i{\vct q}.A^{\dagger}{\vct x}}\,d{\vct x}\\
&=\int\psi^{\dagger}({\vct x}){\mathcal M}_1(0,{\vct q})\psi({\vct x})e^{i{\vct q}.{\vct x}}\,d{\vct x}\\
&=F({\vct q}),
\end{align*}
for all $A\in O(3)$ where $R\in\mbox{Rotations}\subset K$ is such that
\[ \Lambda(R) = \left(
\begin{array}{cc}
1&0\\
0&A
\end{array}\right), \]
and we have used the rotational covariance of ${\mathcal M}_1$ and Theorem~\ref{theorem:Dirac}.
$\Box$

Therefore we have
\begin{align*}
\Delta E&=\omega(4\pi)\int_{s=0}^{\infty}F(0,0,s)s^2\,ds\\
&=4\pi\omega\int_{s=0}^{\infty}\int_{{\bf R}^3}\psi^{\dagger}({\vct x}){\mathcal M}_1(0,0,0,s)\psi({\vct x})e^{i(0,0,s).{\vct x}}s^2\,d{\vct x}\,ds\\
&=-4\pi\omega\int_{s=0}^{\infty}\int_{{\bf R}^3}\psi^{\dagger}({\vct x})N(s)\psi({\vct x})e^{i(0,0,s).{\vct x}}\,d{\vct x}\,ds\\
&=-(4\pi)(2\pi)\omega\int_{s=0}^{\infty}\int_{r=0}^{\infty}\int_{\theta=0}^{\pi}\psi^{\dagger}(r)N(s)\psi(r)e^{isr\cos(\theta)}r^2\sin(\theta)\,d\theta\,dr\,ds\\
&=(4\pi)(2\pi)^{-2}\int_{s=0}^{\infty}\int_{r=0}^{\infty}\psi^{\dagger}(r)N(s)\psi(r)\frac{1}{irs}(e^{isr}-e^{-isr})r^2\,dr\,ds,
\end{align*}
where
\begin{equation}
N(s)={\mathcal N}((0,0,0,\frac{s}{2})).
\end{equation}
Thus writing
\begin{align*}
\Delta V(r)&=(2\pi)^{-2}\int_{s=0}^{\infty} N(s)\frac{1}{isr}(e^{isr}-e^{-isr})\,ds\\
&=(2\pi)^{-2}\frac{1}{ir}\int_{s=0}^{\infty}\frac{N(s)}{s}(e^{isr}-e^{-isr})\,ds,
\end{align*}
we have
\begin{equation}
\Delta E=4\pi\int_{r=0}^{\infty}\psi^{\dagger}(r)\Delta V(r)\psi(r)r^2dr.
\end{equation}
(It can be shown using Fubini's theorem that the order of the integrations can be interchanged.)

Note that $\Delta V:(0,\infty)\rightarrow{\bf C}^{4\times4}$ is a complex matrix valued potential function. We would like to analytically continue $\Delta V$ to the upper imaginary axis of the complex plane, since, for this bound state problem, we are considering spacelike points in Minkowski space for which $x^2<0$ corresponding to pure imaginary $r$ 

\begin{figure} 
\centering
\includegraphics[width=15cm]{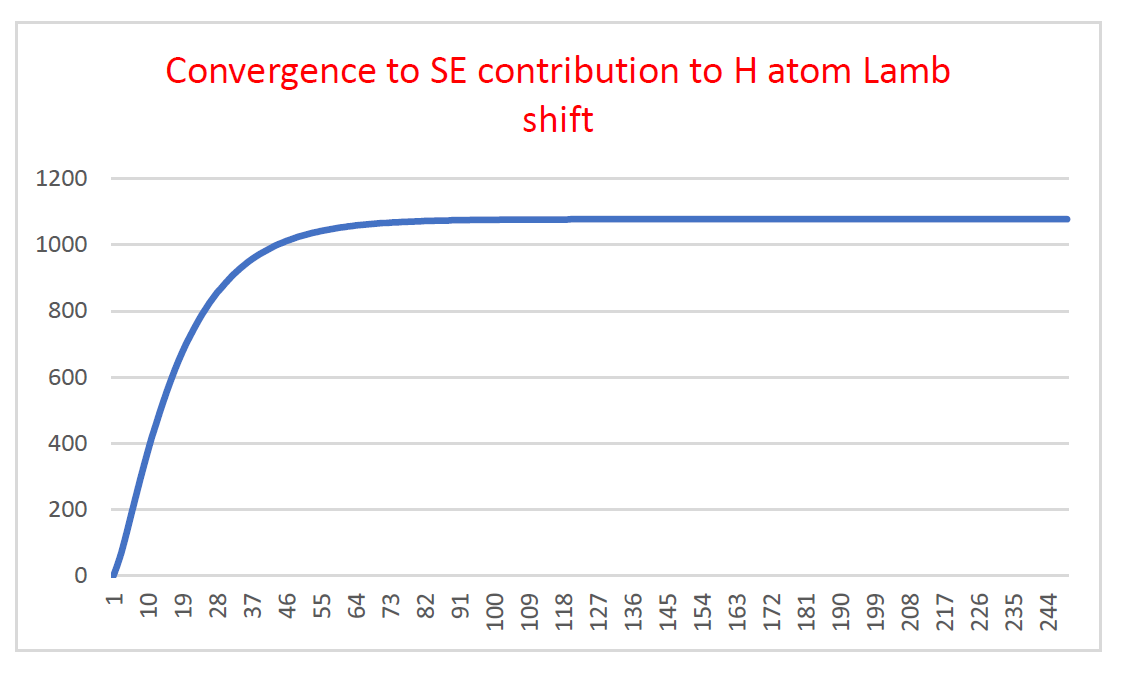}
\caption{Convergence to SE contribution to the Lamb shift for the H atom, MHz vs. iteration} \label{fig:Fig1}
\end{figure}

Applying the transform ${\mathcal T}$ defined in Section~\ref{section:Uehling_contribution} with 
\begin{equation}
F(r)=\int_{s=0}^{\infty}f(s)(e^{-sr}-e^{-isr})\,ds,
\end{equation}
and
\begin{equation}
f(s)=\frac{N(s)}{s},
\end{equation}
we obtain
\begin{align*}
\Delta V(r)&=(2\pi)^{-2}\frac{1}{ir}F(r)\\
&\rightarrow(2\pi)^{-2}\frac{1}{ir}({\mathcal T}(F))(r)\\
&=(2\pi)^{-2}\frac{1}{ir}i\int_{s=0}^{\infty}\frac{N(s)}{s}e^{-sr}\,ds\\
&=(2\pi)^{-2}\frac{1}{r}\int_{s=m}^{\infty} \frac{N(s)}{s}e^{-sr}\,ds.
\end{align*}
We call this the SE potential function and it is analogous to the Uehling potential function. The code for a C++ program to compute the SE contribution to the Lamb shift using the SE potential function is given in the Appendix. The output of the program is shown in Figure~\ref{fig:Fig1}. The output converges to 1077 MHz whereas the accepted value for the SE contribution to the Lamb shift is 1085 MHz \cite{39}. Thus the computed value differs by 0.7\% from the accepted value. However this is not surprising since the program carries out more than 250,000,000 iterations which would be associated with considerable computational error. Accuracy could be increased by using the reduced mass of the electron-proton system rather than the electron mass and also by using the relativistic Dirac H atom 2s eigenfunction rather than the non-relativistic Schr\"{o}dinger eigenfunction.

A great deal of research was done many years ago e.g. \cite{38,41}  on optimizing the numerical computation of the Uehling contribution to the Lamb shift on the basis of the Uehling potential function. One may envisage carrying out similar optimization for the numerical calculation of the electron self-energy contribution to the Lamb shift on the basis of the SE potential function.

Moreover the existence of the simple form of the SE potential obtained without the need for renormalization may simplify many computations in QFT involving multiple loops which are often hampered and complicated by nested divergences e.g. \cite{42,43}. 

\section*{Acknowledgements}

The author is grateful to Christopher Chantler for very helpful discussions and to Vladimir Yerokhin and Randolph Pohl for very helpful comments.

\section*{Appendix: C++ code to compute the electron self-energy contribution to the Lamb shift}

\begin{verbatim}

// self_energy.cpp : This file contains the 'main' function. 
// Program execution begins and ends there.
//

#include <iostream>
#include <fstream>
#include <math.h>

class complex
{
public:
    double real;
    double imaginary;
    complex()
        // constructor
    {
        real = 0.0;
        imaginary = 0.0;
    }
};
//#include "complex.h"

void make_gamma_matrices();
void make_unit();
void create_arrays();
void sum(complex***, complex***, complex***);
void prod(complex***, complex***, complex***);
void vec_prod(complex***, complex**, complex**);
void S(double*, complex***);
void Sigma(double*, complex***);
void compute_N(double*);
double psi(double);
void matrix_dagger_4(complex***, complex***);
void compute_psi(double, complex**);
double psi(double);
void vec_hermitian_prod(complex**, complex**, complex*);

complex**** gamma;
complex*** unit;

const double pi = 4.0 * atan(1.0);

complex*** mat_1, *** mat_2, *** mat_3, *** mat_4, *** mat_5, *** mat_6;
complex*** N;
complex* z4;
complex* Delta_V;
double* p;
double zeta;
complex* Delta_E;
complex** cx_vec, ** cx_vec_1;

const double m_electron = 9.10938356e-31; // electron mass in Kg mks
const double c = 2.99792458e8;  // speed of light m/s mks
const double e = 1.6021766208e-19; // electron charge in Coulombs mks
const double h = 6.626070040e-34; // Planck constant mks
const double h_bar = h / (2.0 * pi);
const double epsilon_0 = 8.854187817e-12; // permittivity of free space mks
const double e1 = e / sqrt(epsilon_0); // electron charge in rationalized units
//const double e1 = e/sqrt(4*pi*epsilon_0); // electron charge in Gaussian units
const double alpha = e1 * e1 / (4 * pi * h_bar * c); // fine structure constant
double m = m_electron * c * c / e; // electron mass in natural units
double a_0_mks = 4.0 * pi * h_bar * h_bar / (m * e1 * e1);
// Bohr radius of the Hydrogen atom in meters
double a_0 = 1.0 / (m * alpha);
// a_0 in natural units eV^{-1}
double factor = e / h; // factor for converting from eV to Hz

double Lambda_r = 5.0 * a_0;
double delta_r = (Lambda_r) / 10000;
int N_s = 100000;
double Lambda_s = 50.0 * m;
double delta_s = (Lambda_s) / N_s;
double delta = delta_r * delta_s;

int main()
{
    std::ofstream outFile("out.txt");
    create_arrays();
    make_gamma_matrices();
    make_unit();

    std::cout << "electron mass = " << m << " eV\n";
    std::cout << "electron mass in Kg = " << m_electron << " Kg\n";
    std::cout << "Inverse fine structure constant 1/alpha = " << 1 / alpha << "\n";
    std::cout << "Bohr radius of hydrogen atom = "
        << a_0_mks << " m\n";
    std::cout << "Bohr radius in natural units = " << a_0
        << " eV^{-1}\n\n";

    Delta_E->real = 0.0;
    Delta_E->imaginary = 0.0;
    int i, j;
    double answer = 0.0;
    for (i = 1;i<250; i++)
    {
        double eV_r, eV_i, MHz_r, MHz_i;
        eV_r = -Delta_E->real * 4.0 * pi * alpha * 4.0 * pi * delta /
                 (2.0 * pi * 2.0 * pi);
        eV_i = -Delta_E->imaginary * 4.0 * pi * alpha * 4.0 * pi * delta /
                (2.0 * pi * 2.0 * pi);
        MHz_r = eV_r * factor / (1.0e6);
        MHz_i = eV_i * factor / (1.0e6);
        std::cout << "for i = " << i << " answer = ("
            << eV_r << "," << eV_i << ") eV = ("
            << MHz_r << "," << MHz_i << ") MHz\n";
        outFile << MHz_r << "\n";
        double r = i * delta_r;
        for (j = 1; j < N_s; j++)
        {
            double s = m + j * delta_s;
            p[0] = 0.0;
            p[1] = 0.0;
            p[2] = 0.0;
            p[3] = s / 2.0;
            zeta = s / 2.0;
            compute_N(p);
            compute_psi(r, cx_vec);
            vec_prod(N, cx_vec, cx_vec_1);
            vec_hermitian_prod(cx_vec, cx_vec_1, z4);
            double v_r = exp(-r * s) * r * z4->real / s;
            double v_i = exp(-r * s) * r * z4->imaginary / s;
            Delta_E->real += v_r;
            Delta_E->imaginary += v_i;
        }
    }
    return 0;
}

void compute_N(double* q_vec)
{
    int i, j, k;
    for (i = 0; i < 4; i++)
        for (j = 0; j < 4; j++)
        {
            mat_1[i][j]->real = (q_vec[0] * gamma[0][i][j]->real +
                q_vec[1] * gamma[1][i][j]->real +
                q_vec[2] * gamma[2][i][j]->real +
                q_vec[3] * gamma[3][i][j]->real) / (m);
            mat_1[i][j]->imaginary = (q_vec[0] * gamma[0][i][j]->imaginary +
                q_vec[1] * gamma[1][i][j]->imaginary +
                q_vec[2] * gamma[2][i][j]->imaginary +
                q_vec[3] * gamma[3][i][j]->imaginary) / (m);
            mat_2[i][j]->real = -mat_1[i][j]->real;
            mat_2[i][j]->imaginary = -mat_1[i][j]->imaginary;
        }
    sum(mat_1, unit, mat_3);
    Sigma(q_vec, mat_4);
    S(q_vec, mat_5);
    prod(mat_4, mat_5, mat_6);
    prod(mat_3, mat_6, mat_4);
    sum(mat_2, unit, mat_5);
    prod(gamma[0], mat_5, mat_6);
    prod(mat_4, mat_6, N);
}

void S(double* p_vec, complex*** answer)
// density for the electron propagator S
{
    int i, j;
    double v = zeta * zeta - m * m;
    if (v != 0.0)
    {
        for (i = 0; i < 4; i++)
            for (j = 0; j < 4; j++)
            {
                answer[i][j]->real = (
                    p_vec[0] * gamma[0][i][j]->real +
                    p_vec[1] * gamma[1][i][j]->real +
                    p_vec[2] * gamma[2][i][j]->real +
                    p_vec[3] * gamma[3][i][j]->real +
                    m * unit[i][j]->real) / v;
                answer[i][j]->imaginary = (
                    p_vec[0] * gamma[0][i][j]->imaginary +
                    p_vec[1] * gamma[1][i][j]->imaginary +
                    p_vec[2] * gamma[2][i][j]->imaginary +
                    p_vec[3] * gamma[3][i][j]->imaginary +
                    m * unit[i][j]->imaginary) / v;
            }
    }
}

void Sigma(double* p_vec, complex*** answer)
// density for the electron self-energy kernel Sigma
{
    int i, j;
    double b = zeta;
    double Z_1 = (b * b - m * m) / (2 * b);
    double Z_1_prime = (b * b + m * m) / (2 * b * b);
    double o_Z = sqrt(Z_1 * Z_1 + m * m);
    for (i = 0; i < 4; i++)
        for (j = 0; j < 4; j++)
        {
            answer[i][j]->real = (4.0 * m * unit[i][j]->real - 2.0 * o_Z *
                gamma[0][i][j]->real) * Z_1 * Z_1_prime / (4.0 * pi * o_Z);
            answer[i][j]->real *= 4.0 * pi * alpha;
            answer[i][j]->imaginary = (-2.0 * o_Z * gamma[0][i][j]->imaginary) *
                Z_1 * Z_1_prime / (4.0 * pi * o_Z);
            answer[i][j]->imaginary *= 4.0 * pi * alpha;

        }
 }

void matrix_dagger_4(complex*** m1, complex*** m2)
{
    int i, j;
    for (i = 0; i < 4; i++)
        for (j = 0; j < 4; j++)
        {
            m2[i][j]->real = m1[j][i]->real;
            m2[i][j]->imaginary = -m1[j][i]->imaginary;
        }
}

void make_gamma_matrices()
{
    gamma = new complex * **[4];
    int i, j, k;
    for (k = 0; k < 4; k++)
    {
        gamma[k] = new complex * *[4];
        for (i = 0; i < 4; i++)
        {
            gamma[k][i] = new complex * [4];
            for (j = 0; j < 4; j++)
                gamma[k][i][j] = new complex;
        }
        for (i = 0; i < 4; i++)
            for (j = 0; j < 4; j++)
            {
                gamma[k][i][j]->real = 0.0;
                gamma[k][i][j]->imaginary = 0.0;
            }
    }
    // Dirac
    gamma[0][0][0]->real = 1.0;
    gamma[0][1][1]->real = 1.0;
    gamma[0][2][2]->real = -1.0;
    gamma[0][3][3]->real = -1.0;

   // make_gamma[1]
    gamma[1][0][3]->real = 1.0;
    gamma[1][1][2]->real = 1.0;
    gamma[1][2][1]->real = -1.0;
    gamma[1][3][0]->real = -1.0;
    // make_gamma[2]
    gamma[2][0][3]->imaginary = -1.0;
    gamma[2][1][2]->imaginary = 1.0;
    gamma[2][2][1]->imaginary = 1.0;
    gamma[2][3][0]->imaginary = -1.0;
    // make_gamma[3]
    gamma[3][0][2]->real = 1.0;
    gamma[3][1][3]->real = -1.0;
    gamma[3][2][0]->real = -1.0;
    gamma[3][3][1]->real = 1.0;
}

void make_unit()
{
    int i, j;
    unit = new complex * *[4];
    for (i = 0; i < 4; i++)
    {
        unit[i] = new complex * [4];
        for (j = 0; j < 4; j++)
            unit[i][j] = new complex;
    }
    for (i = 0; i < 4; i++)
        for (j = 0; j < 4; j++)
        {
            if (i == j)
            {
                unit[i][j]->real = 1.0;
                unit[i][j]->imaginary = 0.0;
            }
            else

            {
                unit[i][j]->real = 0.0;
                unit[i][j]->imaginary = 0.0;
            }
        }
}

void create_arrays()
{
    int i, j;
    // create 4x4 mat matrices

    mat_1 = new complex * *[4];
    for (i = 0; i < 4; i++) mat_1[i] = new complex * [4];
    for (i = 0; i < 4; i++)
        for (j = 0; j < 4; j++)
            mat_1[i][j] = new complex;

    mat_2 = new complex * *[4];
    for (i = 0; i < 4; i++) mat_2[i] = new complex * [4];
    for (i = 0; i < 4; i++)
        for (j = 0; j < 4; j++)
            mat_2[i][j] = new complex;

    mat_3 = new complex * *[4];
    for (i = 0; i < 4; i++) mat_3[i] = new complex * [4];
    for (i = 0; i < 4; i++)
        for (j = 0; j < 4; j++)
            mat_3[i][j] = new complex;

    mat_4 = new complex * *[4];
    for (i = 0; i < 4; i++) mat_4[i] = new complex * [4];
    for (i = 0; i < 4; i++)
        for (j = 0; j < 4; j++)
            mat_4[i][j] = new complex;

    mat_5 = new complex * *[4];
    for (i = 0; i < 4; i++) mat_5[i] = new complex * [4];
    for (i = 0; i < 4; i++)
        for (j = 0; j < 4; j++)
            mat_5[i][j] = new complex;

    mat_6 = new complex * *[4];
    for (i = 0; i < 4; i++) mat_6[i] = new complex * [4];
    for (i = 0; i < 4; i++)
        for (j = 0; j < 4; j++)
            mat_6[i][j] = new complex;

    // create N
    N = new complex * *[4];
    for (i = 0; i < 4; i++) N[i] = new complex * [4];
    for (i = 0; i < 4; i++)
        for (j = 0; j < 4; j++)
            N[i][j] = new complex;

    z4 = new complex;
    Delta_E = new complex;
    p = new double[4];

    // create complex vectors

    cx_vec = new complex * [4];
    for (i = 0; i < 4; i++)
        cx_vec[i] = new complex;
    cx_vec_1 = new complex * [4];
    for (i = 0; i < 4; i++)
    {
        cx_vec_1[i] = new complex;
    }
}

void sum(complex*** M_1, complex*** M_2, complex*** answer)
{
    // form the sum of two complex matrices
    int i, j;
    for (i = 0; i < 4; i++)
        for (j = 0; j < 4; j++)
        {
            answer[i][j]->real = M_1[i][j]->real + M_2[i][j]->real;
            answer[i][j]->imaginary = M_1[i][j]->imaginary + M_2[i][j]->imaginary;
        }
}

void prod(complex*** M_1, complex*** M_2, complex*** answer)
{
    // form the product of two complex matrices
    int i, j, k;
    for (i = 0; i < 4; i++)
        for (j = 0; j < 4; j++)
        {
            answer[i][j]->real = 0.0;
            answer[i][j]->imaginary = 0.0;
            for (k = 0; k < 4; k++)
            {
                answer[i][j]->real += M_1[i][k]->real * M_2[k][j]->real -
                    M_1[i][k]->imaginary * M_2[k][j]->imaginary;
                answer[i][j]->imaginary += M_1[i][k]->imaginary * M_2[k][j]->real +
                    M_1[i][k]->real * M_2[k][j]->imaginary;
            }
        }
}

void vec_prod(complex*** m1, complex** p_vec, complex** answer)
{
    int i, j;
    for (i = 0; i < 4; i++)
    {
        answer[i]->real = 0.0;
        answer[i]->imaginary = 0.0;
        for (j = 0; j < 4; j++)
        {
            answer[i]->real += m1[i][j]->real * p_vec[j]->real - 
                m1[i][j]->imaginary *p_vec[j]->imaginary;
            answer[i]->imaginary += m1[i][j]->imaginary * p_vec[j]->real + 
               m1[i][j]->real *p_vec[j]->imaginary;
        }
    }
}

void vec_hermitian_prod(complex** cx_vec_1, complex** cx_vec_2, complex* z)
{
    z->real = 0.0;
    z->imaginary = 0.0;
    int i;
    for (i = 0; i < 4; i++)
    {
        z->real += cx_vec_1[i]->real * cx_vec_2[i]->real + 
            cx_vec_1[i]->imaginary * cx_vec_2[i]->imaginary;
        z->imaginary += cx_vec_1[i]->real * cx_vec_2[i]->imaginary - 
            cx_vec_1[i]->imaginary * cx_vec_2[i]->real;
    }
}

double psi(double r)
{
    // Hydrogen atom wave function for 2s orbital
    double answer;
    double v = r / (2.0 * a_0);
    answer = (2.0 - r / a_0) * exp(-v);
    answer /= (4.0 * sqrt(2.0 * pi) * a_0 * sqrt(a_0));
    return(answer);
}

void compute_psi(double r, complex** psi_Dirac)
{
// make Dirac wave function with NR Schroedinger wave function as first component

    psi_Dirac[0]->real = psi(r);
    psi_Dirac[0]->imaginary = 0.0;
    psi_Dirac[1]->real = 0.0;
    psi_Dirac[1]->imaginary = 0.0;
    psi_Dirac[2]->real = 0.0;
    psi_Dirac[2]->imaginary = 0.0;
    psi_Dirac[3]->real = 0.0;
    psi_Dirac[3]->imaginary = 0.0;
}

\end{verbatim}


\begin{thebibliography}{99}

\bibitem{60} Lamb, W. E., Retherford, R. C., \textit{The fine structure of hydrogen atoms by a microwave method}, Physical Review 72(3), 241-243, 1947.
\bibitem{51} Kroll, N. M., Lamb, W. E. Jnr., \textit{On the self-energy of a bound electron}, Physical Review 75(3), 388-398, 1949.
\bibitem{45} Bethe, H. A., \textit{The electromagnetic shift of energy levels}, Physical Review 72(4), 339-341, 1947.
\bibitem{50} Feynman, R. P., \textit{Space-time appxoach to quantum electrodynamics}, Physical Review 76(6), 769-789, 1949.
\bibitem{61} Schwinger, J., \textit{Quantum electrodynamics .2. Vacuum polarization and self-energy}, Physical Review 75(4), 651-679, 1949.
\bibitem{46} Dyson, F. J., \textit{The electromagnetic shift of energy levels}, Physical Review 73(6), 617-626, 1948.
\bibitem{39} V. A. Yerokhin, K. Pachucki and V. Patk\'{o}$\chp{\mbox{s}}$, \textit{Theory of the Lamb Shift in Hydrogen and Light Hydrogen-Like Ions}, Ann. Phys. (Berlin) 2019, 531, 1800324.
\bibitem{44} Baranger, M., Bethe, H. A., Feynman, R. P., \textit{Relativistic correction to the Lamb shift}, Physical Review 92(2), 1953, 482-501.
\bibitem{52} Jentschura, U. D., Mohr, P. J., Soff, G., \textit{Calculation of the electron self-energy for low nuclear charge}, Physical Review Letters 82(1), 53-56, 1998.
\bibitem{47} Eides, M. I., Grotch, H. and Shelyuto, V., \textit{Theory of light hydrogenlike atoms}, Physics Reports 342, 63-261, 2001.
\bibitem{48} J. Zamastil and V. Patk\'{o}\v{s}, \textit{Self-energy of an electron bound in a Coulomb field}, Physical Review A 88, Article Number: 032501, 2013.
\bibitem{53} Yerokhin, V. A., Shabaev, V. M., \textit{Lamb shift of n=1 and n=2 states of hydrogen-like atoms, $1\le Z\le110$}, J. Phys. Chem. Ref Data 44, 033103, 2015.
\bibitem{54}Yerokhin, V. A., Harman, Z., \textit{One loop electron self-energy for the bound electron g factor}, Phys. Rev. A 95(6), Article Number: 060501, 2017.
\bibitem{62} Uehling, E. A., \textit{Polarization effects in the positron theory}, Phys. Rev. 48(1), 1935, 55-63.
\bibitem{40} P. Indelicato, P. J. Mohr and J. Sapirstein, \textit{Coordinate-space approach to vacuum polarization}, Physical Review A 89, 042121, 2014.
\bibitem {36} Mashford, J., \textit{Divergence free quantum field theory using a spectral calculus of Lorentz invariant measures}, arXiv:1803.05732, 2018. 
\bibitem{37} Mashford, J., \textit{An introduction to spectral regularization for quantum field theory}, to appear in Proceedings of the XIII International Workshop on Lie Theory and its Applications in Physics (Varna, Bulgaria, June 2019), Springer Proceedings in Mathematics and Statistics, Vol. 335, ed. V. Dobrev, Springer, Heidelberg-Tokyo, 2020.
\bibitem{5} Mashford, J., \textit{An approach to classical quantum field theory based on the geometry of locally conformally flat space-time.} Advances in 
Mathematical Physics, https://doi.org/10.1155/2017/8070462, 2017.
\bibitem{70} Weinberg, S., \textit{The Quantum Theory of Fields}, Volume I, Cambridge University Press, 2005.
\bibitem{38} L. W. Fullerton, and G. A. Rinker Jr., \textit{Accurate and efficient methods for the evaluation of vacuum-polarization potentials of order $Z\alpha$ and $Z\alpha^2$}, Physical Review A, 13(3), 1976, 1283-1287.
\bibitem{22} Mashford J., ``Second quantized quantum field theory based on invariance properties of locally conformally flat space-times",  arXiv:1709.09226, 2017.
\bibitem{41} Huang, K.-N., \textit{Calculation of the vacuum-polarization potential}, Phys. Rev. A 14(4), 1311-1318, 1976. 
\bibitem{42} Bossard G., Kleinschmidt, A., \textit{Cancellation of divergences up to three loops in exceptional field theory}, Journal of High Energy Physics 3, Article Number 100, 2018.
\bibitem{43}  Anirban, B., \textit{Non-analytic terms from nested divergences in maximal supergravity}, Classical and Quantum Gravity 13(14), Article Number 145007, 2016.

\end{thebibliography}
\end{document}